\documentclass{elsarticle}
\makeatletter
\def\ps@pprintTitle{%
 \let\@oddhead\@empty
 \let\@evenhead\@empty
 \def\@oddfoot{\centerline{\thepage}}%
 \let\@evenfoot\@oddfoot}
\makeatother
\usepackage{amsmath,amssymb,amsfonts,theorem}
\usepackage[width=15cm,height=23cm]{geometry}
\usepackage{graphicx}
\usepackage{array,latexsym,amsmath,amsfonts,amssymb,mathrsfs}
\usepackage{psfrag,graphicx,epsfig,color}
\usepackage{changebar}
\usepackage{fancyhdr}
\usepackage{ulem}
\usepackage{hyperref}
\usepackage{xcolor}
\usepackage[utf8]{inputenc}
\usepackage[T1]{fontenc}
\usepackage[ruled,vlined]{algorithm2e}
\newcommand{\algoref}[1]{Algorithm~\ref{#1}}

\newcommand{\sM}{\begin{array}{ccccccccc}}
\newcommand{\eM}{\end{array}}

\renewcommand{\d}{{\,\rm  d}}

\newcommand{\pd}[2]{\displaystyle\frac{\displaystyle\partial #1}{\displaystyle\partial #2}}

\newcommand{\lb}{\left(}
\newcommand{\rb}{\right)}

\newcommand{\sv}{\lb\begin{array}{ccccccccccccccccc}}
\newcommand{\sV}{\left[\begin{array}{ccccccccccccccccc}}
\newcommand{\eV}{\end{array}\right]}
\newcommand{\ev}{\end{array}\rb}

\newcommand{\fempty}[1]{{}}

\newcommand{\sty}[1]{\mbox{\boldmath $#1$}}
\newcommand{\styy}[1]{{\mathbb{#1}}}

\newcommand{\fe}{\sty{ e}}

\newcommand{\fg}{\sty{ g}}

\newcommand{\fn}{\sty{ n}}

\newcommand{\fp}{\sty{ p}}
\newcommand{\fq}{\sty{ q}}
\newcommand{\fr}{\sty{ r}}

\newcommand{\fu}{\sty{ u}}

\newcommand{\fw}{\sty{ w}}
\newcommand{\fx}{\sty{ x}}
\newcommand{\fy}{\sty{ y}}

\newcommand{\fC}{\sty{ C}}

\newcommand{\fG}{\sty{ G}}

\newcommand{\fI}{\sty{ I}}
\newcommand{\fJ}{\sty{ J}}
\newcommand{\fK}{\sty{ K}}

\newcommand{\fM}{\sty{ M}}
\newcommand{\fN}{\sty{ N}}
\newcommand{\fO}{\sty{ 0}}
\newcommand{\fP}{\sty{ P}}

\newcommand{\fS}{\sty{ S}}
\newcommand{\fT}{\sty{ T}}
\newcommand{\fU}{\sty{ U}}
\newcommand{\fV}{\sty{ V}}
\newcommand{\fW}{\sty{ W}}
\newcommand{\fX}{\sty{ X}}

\newcommand{\fZ}{\sty{ Z}}

\newcommand{\ffE}{\styy{ E}}

\newcommand{\ffM}{\styy{ M}}
\newcommand{\ffN}{\styy{ N}}

\newcommand{\ffR}{\styy{ R}}

\newcommand{\flambda}{\mbox{\boldmath $\lambda$}}

\newcommand{\fgamma}{\mbox{\boldmath $\gamma $}}

\newcommand{\cF}{{\cal F}}

\newcommand{\cH}{{\cal H}}

\newcommand{\cP}{{\cal P}}

\newcommand{\cU}{{\cal U}}
\newcommand{\cV}{{\cal V}}

\newcommand{\cY}{{\cal Y}}

\newcommand{\ba}{^{( \alpha )}}

\newcommand{\bi}{^{( i )}}
\newcommand{\bj}{^{( j )}}

\title{An algorithmic comparison of the Hyper-Reduction and the Discrete
Empirical Interpolation Method for a nonlinear thermal problem}

\author{Felix Fritzen$^{1}$, Bernard Haasdonk$^{2}$, David Ryckelynck$^{3}$ and Sebastian Sch\"ops$^{4}$\\[2em]
\small
$^{1}$ University of Stuttgart, EMMA -- Efficient Methods for Mechanical Analysis, Institute of Applied Mechanics (CE), Pfaffenwaldring 7, 70569 Stuttgart, Germany; felix.fritzen@mechbau.uni-stuttgart.de\\[0.5em]
$^{2}$ University of Stuttgart, Institute of Applied Analysis and Numerical Simulation, Pfaffenwaldring 57, 70569 Stuttgart, Germany; haasdonk@mathematik.uni-stuttgart.de\\[0.5em]
$^{3}$ MINES ParisTech, PSL Research University, MAT-Centre des mat\'eriaux, CNRS UMR 7633, BP 87, 91003 Evry, France; david.ryckelynck@mines-paristech.fr\\[0.5em]
$^{4}$ Technical University of Darmstadt, Graduate School of Computational Engineering, Dolivostra\ss{}e 15, 64293 Darmstadt, Germany; schoeps@temf.tu-darmstadt.de
}

\begin{document}
\biboptions{numbers}
\begin{abstract}
	A novel algorithmic discussion of the methodological and numerical differences of competing parametric model reduction techniques for nonlinear problems are presented. First, the Galerkin reduced basis (RB) formulation is presented which fails at providing significant gains with respect to the computational efficiency for nonlinear problems. Renown methods for the reduction of the computing time of nonlinear reduced order models are the Hyper-Reduction and the (Discrete) Empirical Interpolation Method (EIM, DEIM). An algorithmic description and a methodological comparison of both methods are provided. The accuracy of the predictions of the hyper-reduced model and the (D)EIM in comparison to the Galerkin RB is investigated.
All three approaches are applied to a simple uncertainty quantification of a planar nonlinear thermal conduction problem. The results are compared to computationally intense finite element simulations.
\end{abstract}

\maketitle

\section{Introduction}
Numerical models in engineering or natural sciences are getting more and more complex, may be nonlinear 
and depending on unknown or controllable design-parameters. Simultaneously, simulation settings increasingly 
move from single-forward simulations to higher-level simulation scenarios. For example optimization and statistical  investigations require multiple solves, interactive applications require real-time simulation response, or slim-computing environments, e.g. simple controllers, require rapid and memory-saving models. 
For such applications the field of model reduction has gained increasing attention during 
the last decade. The goal is an acceleration of a given numerical model based on construction of a 
low-dimensional approximate surrogate model, the so called reduced order model. Due to the reduced dimension, the 
computation should ideally be rapid, hence be applicable for the mentioned multi-query, real-time 
or slim-computing simulation scenarios. 
Well-known techniques for linear problems comprise Proper Orthogonal Decomposition (POD) \cite{Volkwein2017,Volkwein2001}, 
 control-theoretic approaches such as Balanced Truncation, Moment Matching or 
 Hankel-norm approximation \cite{An05}.
For parametric problems, certified Reduced Basis (RB) methods have been developed \cite{PR07,Haasdonk2017}.
Nonlinear problems pose additional challenges. In particular there exists a well-known drawback with 
POD, which is that a high-dimensional reconstruction of the reduced solution is required for 
each evaluation of the nonlinearity. 
Some solution techniques exist, which provide a remedy of this problem. Mainly, these approaches are
sampling-based techniques such as Empirical Interpolation 
\cite{BMNP04} and discrete variants \cite{HO08hyp,DHO12,CS09,CS12} 
or Hyper-Reduction \cite{Ryckelynck-2005,Ryckelynck-2009}. Note, that also further approaches exist, such as Gappy-POD \cite{Everson-1995}, Missing 
Point Estimation (MPE) \cite{AstridWeilandEtAl2008} 
or Gauss-Newton with approximated Tensors (GNAT) \cite{CBF11}.
Most of those methods identify a subset of the arguments of the nonlinear function. Then, based solely on 
evaluation of these few components, they construct an approximation of the full solution.
Such nonlinear problems emerge in many applications. For instance the effective behavior of dissipative microstructured materials needs to be predicted by nonlinear homogenization techniques. These imply a multi-query context in the sense of different loadings (and load paths) applied to the reference volume element in order to obtain the related effective mechanical response. Reduced basis methods combining the purely algorithmic gains of the reduced basis with a reformulation of the problem incorporating micromechanics have shown to be efficient for the prediction of the effective material response, for multi-level FE simulations and for nonlinear multiscale topology optimization (e.g., \cite{fritzen2015a,fritzen2016a}).

{In this paper, we analyze and compare two of those efficient methods, namely Discrete Empirical Interpolation (DEIM) and Hyper-Reduction (HR), for the reduction of a non-trivial nonlinear parametric thermal model. The comparison is carried out on the numerical, algorithmical and mathematical level.}

The paper is structured as follows. We introduce the nonlinear parametrized 
thermal model problem in Section \ref{sec:problem}.
Subsequently, in Section \ref{sec:methods} the methods under investigation are introduced and formally compared: 
As a benchmark the POD-Galerkin procedure is formulated, then the DEIM and the hyper-reduction technique. 
The numerical comparison is provided in Section \ref{sec:experiments}. {A classical scenario from uncertainty quantification is demonstrated since model order reduction is particularly interesting for such many-query scenarios}. Finally, we conclude in 
Section \ref{sec:conclusion}.

\subsection{Nomenclature}
In the manuscript the following notation is used: bold face symbols denote vectors (lowercase letters) or matrices (uppercase letters), The spatial gradient operator is denoted by $\nabla \bullet$ and the divergence is expressed by $\nabla \cdot \bullet$. The dependence on spatial coordinates is omitted for simplicity of notation.
\section{Nonlinear reference problem}
\label{sec:problem}
\subsection{Strong formulation}
\begin{figure}[!h]
\centering
\includegraphics[scale=1]{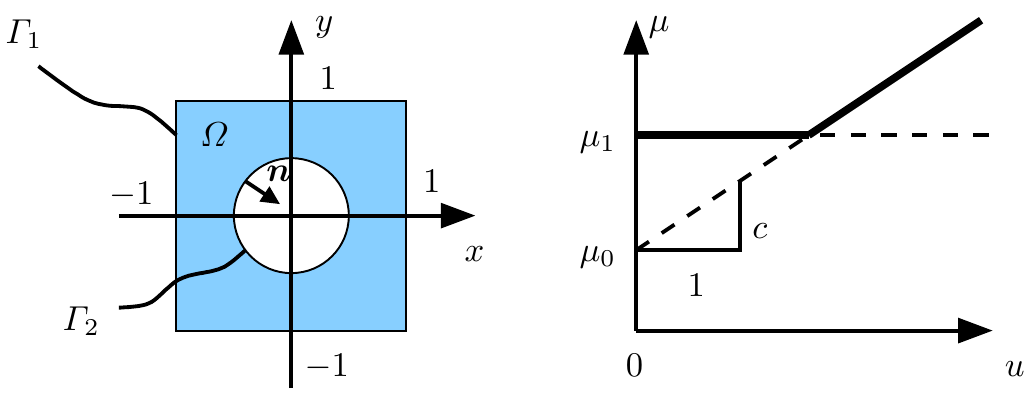}
\caption{Geometry of the planar benchmark problem (left) and nonlinearity of the temperature dependent conductivity $\mu$ (right)}
\label{fig:geometry}
\end{figure}

In the following we consider a stationary planar nonlinear heat conduction problem on a plate with a hole, see 
Fig.~\ref{fig:geometry}. The problem is parametrized by a vector $\fp$. The nonlinearity of the problem is introduced by an 
isotropic Fourier-type heat conductivity~$\mu(u; \fp)$ that depends on the current temperature~$u$ and the parameter vector. 
The Dirichlet and Neumann boundary data are denoted by $u_*$ and $q_*$, respectively. 
Note that in the example $q_*=0$ is considered. However, the formulation of the weak form is considered for the sake of generality.
The corresponding Dirichlet and Neumann boundaries are denoted by $\varGamma_1$ and $\varGamma_2$, respectively.
The strong formulation of the boundary 
value problem is
\begin{align}
 - \nabla \cdot \lb \mu(u; \fp) \, \nabla u \rb & = 0 \ \text{in }\varOmega, &
 u & = u_*(\fp) \ \text{on }\varGamma_1, &
 - \mu(u; \fp) \, \nabla u \cdot \fn & = q_* \ \text{on }\varGamma_2,
 \label{eq:heatproblem}
\end{align}
where $\mu$ is the temperature dependent conductivity.

\subsection{Weak formulation}
The weak form of the heat conduction problem \eqref{eq:heatproblem} is given by
\begin{align}
 a( u, {\delta u}; \fp) - l( {\delta u}; \fp)&=0
 \qquad (\forall {\delta u} \in \cV_0)
\end{align}
where
\begin{align}
 a( u, {\delta u}; \fp) = \intop_\varOmega
 \mu(u; \fp) \, \nabla  u  \cdot \nabla {\delta u} \d \varOmega,
 \quad\text{and}\qquad
 l( {\delta u}; \fp) =- \intop_{\varGamma_2} {\delta u} \, q_* \d \varGamma
\end{align}
with the unknown temperature field $u \in \cV = \cV_0 + \{\bar{u}_*(\fp)\}$ being sought-after.
The function space $\cV_0 := \{u \in \cH^1(\varOmega) : \,  u = 0 \text{ on }\varGamma_1\}$ is 
referred to as space of test functions vanishing on the Dirichlet boundary~$\varGamma_1$ {and $\bar{u}_*(\fp)\in\cH^1(\varOmega)$ is a field defined on the full domain that satisfies the Dirichlet conditions}. 
The Dirichlet conditions on $\varGamma_1$ are assumed to depend on two parameters $g_{\rm x}$ and $g_{\rm y}$ via
\begin{align}
 u_*(\fp) & := g_{\rm x} x + g_{\rm y} y, \quad g_{\rm x}, g_{\rm y} \in [0, 1], \quad x, y \in {\varGamma_1}.
\label{eqn:ustar_definition}
\end{align}
{A trivial choice is to set $\bar{u}_*(\fp)= g_{\rm x} x + g_{\rm y} y$ in the full domain. This particular choice is considered in the remainder of this study.}
While the solution space $\cV$ depends on the parameters~$\fp$ via $\bar{u}_*$, the space of test functions~$\cV_0$ is independent of the parameters~$\fp$. For the heat conductivity $\mu$, an explicit dependence on the temperature via the nonlinear constitutive model
\begin{align}
 \mu(u; \fp) &:= {\rm max} \left\lbrace \mu_1, \mu_0 + c u  \right\rbrace
\end{align}
is assumed (see also Fig.~\ref{fig:geometry}, right). The parameter vector $\fp$ in the present context is
\begin{align}
 \fp & : = [ g_{\rm x}, g_{\rm y}, c, \mu_0, \mu_1 ],
\end{align}
{and attention will be confined to the compact parameter domain}
\begin{align}
  \fp & \in \mathcal{P}:=[0,1] \times [0,1] \times [1, 2] \times \{ 1 \} \times \{0.5\} {\subset \ffR^5}. \label{eq:param:range}
\end{align}

\subsection{Discrete formulation}
\label{sec:discrete}
In order to solve the nonlinear problem \eqref{eq:heatproblem} nodal Finite Elements (FE) are used in a classical Galerkin formulation (e.g., \cite{bathe02,zienkiewicz06}). The space of finite element test functions $\cV^{\rm h}_0\subset\cV_0$ is assumed to be spanned by~$n$ linearly independent and continuous ansatz functions~$\varphi_i$ associated with nodes $(x_i,y_i)$ and $\varphi_i(x_j,y_j)=\delta_{ij}$ where $i,j=1,\ldots,n$. We assume a solution expansion {into a linear field~$\bar{u}_*$ defined over the full domain $\varOmega$ and comply to the parameterized boundary data and a fluctuation term according to}
\begin{align}
 u_h(\fp) & := {\bar{u}_*(\fp)} + \sum_{j=1}^n {w_{h,j}}(\fp) \varphi_j \label{eq:solution-expansion}, &
 {\bar{u}_*(\fp)} & {:= g_{\rm x} x + g_{\rm y} y \quad \forall x,y \in \varOmega} .
\end{align}
{Here, the coefficient vector $\fw(\fp)=\bigl(w_{h,j}(\fp)\bigr)_{j=1}^n \in \ffR^n$ contains the nodal temperature fluctuations which represent the unknowns of the system.
They define the coefficient vector containing the nodal temperatures via
\begin{align}
 \fu (\fw;\fp) &= \fw + \bar{\fu}_*(\fp), 
\end{align}
where $\bar{\fu}_*(\fp)$ is the vector composed of the nodal values of the superimposed linear field $\bar{u}_*(\fp)$. The} discrete nonlinear equations that have to be solved for any given parameter vector~$\fp$ are given by
\begin{align}
 r_i ( {\fw}; \fp )  & :=  a \Big(  {\bar{u}_*} ( \fp ) + \sum_{j=1}^n {w_{h,j}} \varphi_j, \,\varphi_i; \fp \Big) - l(\varphi_i;\fp) = 0  \quad \big( \forall i \in \{1, \dots, n\} \big) .
 \label{eq:residual:i}
\end{align}
{In our specific numerical example the linear functional $l(\varphi_i; \fp)$ is zero due to the chosen homogeneous Neumann conditions. However, the problem \eqref{eq:residual:i} still has a nontrivial solution due to the inhomogeneous Dirichlet data provided through $\bar{u}_*(\fp)$.}
The condition \eqref{eq:residual:i} can be expressed in a compact representation using the vector notation
\begin{align}
 \fr( {\fw}; \fp) & := \big( r_i({\fw};\fp) \big)_{i=1}^n =  \fO. \label{eq:residual:vec}
\end{align}
In order to solve the nonlinear problem~\eqref{eq:residual:vec}, expressed component-wise in~\eqref{eq:residual:i}, the finite 
element method is used to provide the functions for the expansion \eqref{eq:solution-expansion}.
In the following~$\fN$ denotes a row vector of finite element ansatz functions for the temperature, 
and~$\fG$ and $\fG^T$ are the discrete gradient and divergence matrices, respectively. The FE approximation of the temperature~$u$ and its gradient~$\nabla u$ are given by
\begin{align}
 u_{h}(\fw;\fp) &= \fN \fu(\fw;\fp) {= \fN \fw(\fp) + \bar{u}_*( \fp ),} \\
 \nabla u_{h}(\fw; \fp) &= \fG \fu(\fw; \fp) {= \fG \fw(\fp) + \bar{\fg}(\fp),} &
 {\bar{\fg}(\fp)} &{= \sV g_{\rm x} \\ g_{\rm y} \eV}.
\end{align}
Then the exact Jacobian of the finite element system is
\begin{align}
 \fJ_* ( {\fw}; \fp ) &= \intop_\varOmega \mu( u_{h}(\fw;\fp); \fp ) \, \fG^T \fG + \pd{\mu( u_{h}(\fw;\fp); \fp )}{u} \, { \lb \fG^T \fG \fu(\fw;\fp) \rb } \fN \, \d \varOmega.
\end{align}
In the following a symmetric approximation of~$\fJ_*$ given by
\begin{align}
 \fJ (  {\fw}; \fp ) &= \intop_\varOmega \mu( u_{h}  {(\fw; \fp)}; \fp ) \, \fG^T \fG \, \d \varOmega \label{eq:J:approx}
\end{align}
is used. The numerical solution of the nonlinear problem~\eqref{eq:residual:vec} is then obtained by a fixed-point scheme 
using~$\fJ$ as an approximation of the differential stiffness matrix~$\fJ_*$. The resulting iteration scheme is commonly referred 
to as successive substitution (\cite{Kelley_1995aa}, page 66). Note also that $\fJ$ {is well defined for arbitrary thermal fields,} while the exact Jacobian is not defined\footnote{{More precisely, the Jacobian is semi-smooth.}} for the critical temperature~$u_{\rm c}=(\mu_1-\mu_0)/c$ which implies a non-differentiability of the conductivity~$\mu$.

\algoref{algo:fe-solution} illustrates the nonlinear finite element solution procedure for the considered problem. The comment 
lines in the algorithm comprise references to computational costs $c_{\rm reloc}$ 
(computation of $u$ and $\nabla u$), ~$c_{\rm const}$ (for the constitutive model), $c_{\rm rhs}$ (linked to 
the residual computation), $c_{\rm Jac}$ (related to the Jacobian) and $c_{\rm sol}$ (for the linear solver). 
This will become relevant in the comparison section.

\begin{algorithm}[h]
\SetSideCommentRight
\SetNoFillComment
\SetAlgoSkip{medskip}
\LinesNumbered
\SetKwInOut{Input}{Input}
\SetKwInOut{Output}{Output}
\SetArgSty{text}
\caption{finite element solution}
\label{algo:fe-solution}

\Input{parameters $\fp\in\cP$, scatter/gather operator $\fP_e$ of element $e$ and weight $v_i$ at the integration point $\fx_i$ $(i=1, \dots, n_{\rm gp})$}
\Output{ nodal temperature vector $\fu(\fp)$}\vskip2pt
\hrule\vskip2pt
set $\fu^{(0)}={\bar{\fu}}_\ast(\fp)$; $\alpha=0$ \tcp*{initialization}
set $\fr=\fO$, $\fJ=\fO$   \tcp*{reset r.h.s. and Jacobian}
\For(\tcp*[f]{{$n_{\rm el}$: number of elements}}){$e=1, \dots, n_{\rm el}$}{
set $\fr_e=\fO$, $\fJ_e=\fO$ \tcp*{reset element r.h.s. and Jacobian}
\For(\tcp*[f]{{$n_e^{\rm gp}$: number of int. point in element $e$}}){$i=1, \dots, n_e^{\rm gp}$}{
{evaluate the FE matrices $\fN, \fG$ and $\fG_e = \fG \fP_e $ at the current int. point}\tcp*{$c_{\rm reloc}$}
compute $u_h\ba \leftarrow \fN \fu\ba { + \bar{u}_*(\fp)}$ and gradient $\fg_h\ba \leftarrow \fG\fu\ba$ at int. point \tcp*{$c_{\rm reloc}$}
evaluate constitutive model $\mu\ba \leftarrow \mu(u_h\ba; \fp)$ \tcp*{$c_{\rm const}$}
$\fr_e \leftarrow \fr_e + v_i \, \mu\ba \, {\fG_e}^T \fg_h\ba$ \tcp*{$c_{\rm rhs}$}
$\fJ_e \leftarrow \fJ_e + v_i \, \mu\ba \, {\fG_e}^T \fG_e $ \tcp*{$c_{\rm Jac}$}
}
$\fr \leftarrow \fr +  \fP^T_e  \fr_e $; $\fJ \leftarrow \fJ + \fP^T_e \fJ_e \fP_e$ \tcp*{assembly; $c_{\rm rhs}, c_{\rm Jac}$}
}
solve $\fJ\delta\fu\ba = -\fr$ and update nodal temperatures $\fu^{(\alpha+1)}\leftarrow\fu\ba+\delta\fu\ba$; $\alpha\leftarrow\alpha+1$  \tcp*{$c_{\rm sol}$}
converged ($\Vert\delta u_h\ba \Vert_{L^2(\varOmega)}<\epsilon_{\rm max}$)? $\to$ {\bf end}; \quad {\it else:} goto \NlSty{2}
\end{algorithm}

\subsection{Reduced basis ansatz}
In the following an $m$-dimensional basis of global ansatz functions $\psi_k$ is considered in the 
reduced framework, where $m \ll n$ is implicitly asserted in order to obtain an actual model order reduction.
The functions $\psi_k$ define a linear subspace of $\cV_0^{\rm h}$ which can be characterized 
by a matrix $\fV =  (V_{ij}) \in \ffR^{n\times m}$ via
\begin{align}
 \psi_k & : = \sum_{i=1}^n V_{ik} \varphi_i && (k=1, \dots, m).
\end{align}
The reduced coefficient vector ${\fgamma (\fp)=(\gamma_k(\fp))_{k=1}^m\in \ffR^m}$ defines the temperature {fluctations}
in the reduced setting by the affine relation
\begin{align}
\tilde{u} (\fp) = \tilde{u} ( \fgamma(\fp); \fp ) &:= u_* ( \fp ) + \tilde{w}(\fp), &
\tilde{w}(\fp) & := \sum_{k=1}^m \gamma_k { (\fp) }\, \psi_k . \label{eqn:reduced-solution}
\end{align}
The corresponding vector of discrete values~${\tilde{\fu}(\fp) := (\tilde u_{j}(\fp))_{j=1}^n \in \ffR^n}$ of the 
temperature with respect to the basis of $\cV^{\rm h}_0$ is given by the linear 
relation 
$ \tilde{\fu} = \bar{\fu}_*(\fp) + \tilde{\fw}(\fp)$ with the coefficient vector $\tilde{\fw}=\fV \fgamma {(\fp)}=(\tilde{w})_{j=1}^n$ of the reduced fluctuation field. Hence, the reduced fluctuation $\tilde w$ can equivalently be expressed as 
\begin{align}
  \tilde{{w}}(\fp) = \sum_{j=1}^n \tilde w_{j}{ (\fp) } \varphi_j
  = \sum_{j=1}^n \varphi_j \sum_{k=1}^m V_{jk} \gamma_k{ (\fp) }.
\end{align}
In the following the matrix $\fV$ defining the space of the reduced ansatz functions is assumed to 
be given by a snapshot POD~\cite{Sirovich_1987} obtained from full resolution finite element simulations evaluated at 
$s$ different parameter vectors $\fp\bi$ ($i=1, \dots, s$). 
The POD subspace ${\tilde \cV^{\rm h}_0\subset \cV^{\rm h}_0}$ 
is then obtained as best-approximating $m$-dimensional subspace with respect to the $L^2(\varOmega)$-error, i.e. 
\begin{align}
  \tilde{\cV}^{\rm h}_0 &= \underset{\stackrel{\tilde \cV \subset \cV^{\rm h}_0}{\mathrm{dim}(\tilde \cV) = m}}{\rm arg \, min }
  \sum_{i=1}^s \left\lVert {w_{h}}(\fp^{(i)})  - P_{\tilde \cV}{w_{h}}(\fp^{(i)}) \right\rVert^2_{L^2(\varOmega)},
\end{align}
where $P_{\tilde \cV}$ is the orthogonal projection {of the thermal fluctuations} onto the subspace $\tilde \cV$. 
Numerically, a POD-basis of this space can be obtained by a corresponding matrix eigenvalue problem, 
e.g., cf.\ \cite{Jo:02,Volkwein2017,Haasdonk2017}. Here, we use the $L^2(\varOmega)$ inner product for the 
computation of the symmetric snapshot inner product 
matrix
\begin{align}
 {\fC} &= ({C}_{ij})_{i,j=1}^s\in\ffR^{s \times s}, &
  {C}_{ij} & := \intop_\varOmega  {w_{h}}(\fp^{(i)}) {w_{h}}(\fp^{(j)}) \d \varOmega. 
\end{align}
{The entries of $\fC$ are defined by the discrete snapshot solutions $\fw\bi$ via the mass matrix $\fM$ as follows
\begin{align}
 \fM &:= \intop_\varOmega \fN^T \fN \, \d \varOmega, &
 C_{ij} & = \lb\fw\bi\rb^T \fM \lb\fw\bj\rb.
 \end{align}}%
A pseudo-code implementation of the snapshot POD is given by~\algoref{algo:snapshot-pod}.
{Alternatively, a weighted SVD of the snapshot matrix can be used to obtain the same basis.}

\begin{algorithm}[h]
\SetSideCommentRight
\SetNoFillComment
\SetAlgoSkip{medskip}
\LinesNumbered
\SetKwInOut{Input}{Input}
\SetKwInOut{Output}{Output}
\SetArgSty{text}
\caption{snapshot proper orthogonal decomposition (snapshot POD)}
\label{algo:snapshot-pod}

\Input{ $s$ snapshot parameters $\fp\bi \subseteq \cP$, reduced dimension~$m$ {\it (or: threshold $\delta$)} and {mass matrix $\fM$}}
\Output{ reduced basis functions $\psi_k$ $(k=1,\dots,m)$ and coefficient matrix $\fV$ }\vskip2pt
\hrule\vskip2pt

\For{$i=1, \dots, s$}{
solve high-fidelity problem (e.g. via FE solution; \algoref{algo:fe-solution}) for parameter $\fp\bi$ $\to$ ${\fw\bi}$\\
store ${\fw\bi}$ as $i$-th column of the snapshot matrix $\fS$
}
compute snapshot inner product matrix ${\fC := \fS^T\fM\fS}$\\
compute $m$ leading eigenpairs $(\fq_k,{\xi}_k)$ of ${\fC}$ ($k=1, \dots, m$)\\
set $V_{jk}:= \lb {\xi}_k \rb^{-1/2} \sum_{i=1}^s  \lb{\fw\bi}\rb_j \lb \fq_k \rb_i \quad (j=1,\dots,n)$ and $\psi_k=\sum_{j=1}^n \varphi_j V_{jk}$
\end{algorithm}

There are many other techniques to determine reduced projection spaces, e.g., greedy techniques, balanced truncation, Hankel-norm approximation, moment matching and others (e.g., \cite{An05}). 
However, we do not aim at a comparison of projection space techniques, but rather focus on the treatment of the nonlinearities. Therefore, we decide for the POD space as 
common reduced projection space for all subsequent methods.

Note, that in view of \eqref{eqn:ustar_definition} we have parameter separability for the Dirichlet data with two parameter-independent components. The discrete representation of the latter is given by
\begin{align}
{\bar{\fu}_*}(\fp) = g_{\rm x} {\bar{\fu}}_{*,1} + g_{\rm y} {\bar{\fu}}_{*,2}.
\end{align}
Hence, any reduction scheme can still provide an approximation of~$u(\fp)$ via~\eqref{eqn:reduced-solution} which exactly matches the prescribed Dirichlet data. The components of the vectors $\fu_{*,1}$, $\fu_{*,2}$ are chosen as
\begin{align}
  \lb u_{*,1} \rb_{i=1}^n &: = x_i, &
  \lb u_{*,2} \rb_{i=1}^n &: = y_i,
\end{align}
where $x_i$ and $y_i$ denote the coordinates of node number~$i$. Thereby, constant gradients within the structure are also represented exactly.

\section{Sampling-based Reductions}
\label{sec:methods}
\subsection{Galerkin reduced basis approximation}
\label{sec:galerkin}
A classical Galerkin projection on a POD space is used to provide reference solutions for the other reduction methods. In order to solve the weak form of the nonlinear problem for given parameters $g_{\rm x}$, $g_{\rm y}$, a successive substitution is performed in which the local conductivity~$\mu$ is computed using the previous iteration of the temperature~$\tilde{u}\ba := u_\ast(\fp)+\tilde{w}\ba$ with $\tilde{w}\ba=\fN\fV\fgamma\ba$ in the reduced setting, where $\alpha \in \ffN$ is the iteration number and $\fgamma\ba$ is the iterate of the reduced degrees of freedom in iteration~$\alpha$.
As in a classical Galerkin approach, we assume the (reduced) test function
\begin{align}
 \tilde{v} & : = \sum_{i=1}^n \sum_{j=1}^m \varphi_i \, V_{ij} \lambda_j
\end{align}
with arbitrary coefficients $\lambda_j$ ($j=1, \dots, m$) represented by the vector~$\flambda$.
For convenience, we define the conductivity corresponding to the $\alpha$-th iterate {$\fgamma\ba$ of the reduced coefficient vector $\fgamma(\fp)$}
\begin{align}
 \mu\ba & := \mu \lb \tilde{u} ( \fgamma\ba; \fp); \fp \rb .
\end{align}
Here $\fgamma\ba$ can be interpreted as a constant parameter similar to $\fp$ during the subsequent iteration which provides the new iterate $\fgamma^{(\alpha+1)}$ as the solution of
\begin{align}
	a ( \tilde{u}, \tilde{v}; \fp ) & =
		 \sum_{j=1}^m   \lambda_j \,
				\intop_\varOmega \mu\ba \, \nabla \tilde{u}(\fgamma^{(\alpha+1)}; \fp) \cdot \nabla\psi_j \d \varOmega = 0
					\qquad \forall \flambda \in \ffR^m.
	\label{eq:fixed:point}
\end{align}
Rewriting~\eqref{eq:fixed:point} using the approximated Jacobian~$\fJ$ and the residual $\fr$ leads to the linear system
\begin{align}
 \fV^T \fJ( \fV\fgamma^{(\alpha)}; \fp ) \fV ( \fgamma^{(\alpha+1)} - \fgamma^{(\alpha)} ) + \fV^T \fr( \fV \fgamma^{(\alpha)}; \fp ) & = \fO .
 \label{eq:fixed:point:2}
\end{align}
The projection of the residual onto the reduced basis defines the reduced residual
\begin{align}
  \tilde{\fr}\ba &: = \fV^T \fr( \fV \fgamma\ba; \fp ).
\end{align}%
In view of the definition of the reduced basis by $L^2(\varOmega)$-orthogonal POD modes and with $\delta\fgamma\ba:=\fgamma^{(\alpha+1)}-\fgamma\ba$ one obtains
\begin{align}
  \Vert \delta\fgamma\ba \Vert_{l^2} & = \Vert \tilde{u}^{(\alpha+1)} - \tilde{u}\ba \Vert_{L^2(\varOmega)}
  { = \Vert \tilde{w}^{(\alpha+1)} - \tilde{w}\ba \Vert_{L^2(\varOmega)} }.
\end{align}
This gives rise to the simple convergence criterion $\Vert\delta\fgamma\ba\Vert_{l^2}<\epsilon_{\rm max}$, i.e. the iteration should stop upon sufficiently small changes of the temperature field. \algoref{algo:rb-solution} summarizes the online phase of the Galerkin RB method.

\begin{algorithm}[h]
\SetSideCommentRight
\SetNoFillComment
\SetAlgoSkip{medskip}
\LinesNumbered
\SetKwInOut{Input}{Input}
\SetKwInOut{Output}{Output}
\SetArgSty{text}
\caption{Galerkin reduced basis solution (online phase)}
\label{algo:rb-solution}

\Input{ parameters $\fp\in\cP$; reduced temperature matrix $\fT_i {:= \fN(\fx_i) \fV }$, gradient matrix $\fG_i{:= \fG(\fx_i) \fV }$ and weight $v_i$ at the integration point $\fx_i$ $(i=1, \dots, n_{\rm gp})$}

\Output{ reduced vector $\fgamma$ and nodal temperatures $\tilde{\fu}$ {\it (optional)}}\vskip2pt

\hrule\vskip2pt
set $\fgamma^{(0)}=\fO$; $\alpha=0$ \tcp*{initialization}
set $\tilde{\fr}\ba=\fO$, $\tilde{\fJ}\ba=\fO$   \tcp*{reset r.h.s. and Jacobian}
\For({\tcp*[f]{$n_{\rm gp}$: number of int. points in the mesh}}){$i=1, \dots, n_{\rm gp}$}{
compute temperature $\tilde{u}\ba \leftarrow \fT_i\fgamma\ba + u_\ast$ and gradient $\tilde{\fg}\ba \leftarrow \fG_i\fgamma\ba$ \tcp*{$c_{\rm reloc}$}
evaluate constitutive model $\mu\ba \leftarrow \mu(\tilde{u}\ba; \fp)$ \tcp*{$c_{\rm const}$}
$\tilde{\fr}\ba \leftarrow \tilde{\fr}\ba + {v_i} \, \mu\ba \, \fG_i^T \tilde{\fg}\ba$ \tcp*{$c_{\rm rhs}$; direct assembly}
$\tilde{\fJ}\ba \leftarrow \tilde{\fJ}\ba + {v_i} \, \mu\ba \, \fG_i^T \fG_i $ \tcp*{$c_{\rm Jac}$; direct assembly}
}
solve $\tilde{\fJ}\ba\delta\fgamma\ba=-\tilde{\fr}\ba$ and update $\fgamma^{(\alpha+1)}\leftarrow \fgamma\ba +\delta\fgamma\ba$; set $\alpha\leftarrow\alpha+1$ \tcp*{$c_{\rm sol}$}
converged ($\Vert\delta\tilde{u}\ba\Vert_{L^2(\varOmega)}=\Vert\delta\fgamma\ba\Vert_{l^2}<\epsilon_{\rm max}$)? $\to$ {\bf end}; \quad {\it else:} goto \NlSty{2}
\end{algorithm}

The projected system can be interpreted as a finite element method with global, problem specific ansatz functions~$\psi_k$, whereas the classical finite element method uses local and rather general ansatz functions~$\varphi_j$ (e.g., piecewise defined polynomials).

Note that the solution of~\eqref{eq:fixed:point}, \eqref{eq:fixed:point:2} is also a minimizer of the potential
\begin{align}
 \Pi( \fgamma; \fgamma\ba, \fp ) & : =  \frac{1}{2} \intop_\varOmega \mu\ba \, \nabla \tilde{u}(\fgamma; \fp) \cdot \nabla \tilde{u}(\fgamma; \fp) \, \d \varOmega. 
\end{align}
Therefore, variational methods can directly be applied to solve the minimization problem and alternative numerical strategies are available. Such variational schemes are also used, e.g., in the context of solid mechanical problems involving internal variables (e.g., \cite{fritzen2014a}).

The Galerkin RB method with well-chosen reduced basis functions $\psi_k$ (represented via the matrix $\fV$) can replicate the FEM solution to a high accuracy (see Section~\ref{sec:experiments}). It also provides a significant reduction of the memory requirements: instead of $\fu\in\ffR^n$ only $\fgamma\in\ffR^m$ needs to be stored. Despite the significant reduction of the number of unknowns from $n$ to $m$, the Galerkin RB cannot attain substantial accelerations of the nonlinear simulation
due to a computationally expensive assembly procedure with complexity~$\mathcal{O}(n_{\rm gp})$ for the residual vector~$\fr$ and the fixed point operator~$\fJ$ (compare $c_{\rm rhs}$ and $c_{\rm Jac}$ in \algoref{algo:rb-solution} and \algoref{algo:fe-solution}). Here $n_{\rm gp}$ is the number of quadrature points in the mesh. However, if the linear systems are not solved with optimal complexity, e.g., using sparse $LU$ or Cholesky decompositions with at least $\mathcal{O}(n^2)$, then a reduction of complexity can still be achieved. It shall be pointed out that for very large $n$ (i.e. for millions of unknowns) the linear solver usually dominates the overall computational expense. Then the Galerkin RB may provide good accelerations without further modifications.

\medskip

In order to significantly improve on the computational efficiency while maintaining the reduced number of degrees of freedom (and thus the reduced storage requirements), the Hyper-Reduction \cite{Ryckelynck-2005,Ryckelynck-2009} and 
the Discrete Empricial Interpolation Method (DEIM, \cite{HO08hyp,DHO12,CS09}) 
are used. Both methods are specifically designed for the computationally 
efficient approximation of the nonlinearity of PDEs.

\subsection{Discrete Empirical Interpolation Method (DEIM)}

The empirical interpolation method (EIM) was introduced by \cite{BMNP04} to approximate 
parametric or nonlinear functions by separable interpolants. This technique is meanwhile 
standard in the reduced basis methodology for parametric PDEs.  Discrete versions of the EIM 
for instationary problems have been introduced as empirical operator interpolation 
\cite{HOR08,HO08hyp,DHO12} or alternatively (and in some cases equivalently) as discrete empirical 
interpolation (DEIM) \cite{CS09,CS12}. In particular a-posteriori \cite{HOR08,DHO12,WSH14} 
and a-priori \cite{CS12} error control is possible under certain assumptions, see also \cite{Haasdonk2017}.

We present a formulation for the present stationary problem. Instead of approximating a continuous 
field variable, the goal of the discrete versions is to provide an approximation $\tilde \fr$ for 
the vectorial nonlinearity of the nodal residual vector $\fr$ of the form 
\begin{align}
\tilde \fr(\fu;\fp) := \fU (\fP^T \fU)^{-1} \fP^T \fr(\fu;\fp), \label{eqn:EI-definition}
\end{align}
where the columns of $\fU \in \ffR^{n\times M}$ are called {\it collateral reduced basis} 
and $\fP = [\fe_{i_1},\ldots, \fe_{i_M}]\in \ffR^{n\times M}$ is a sampling matrix with interpolation 
indices (also known as {\it magic points}) $i_1,\ldots, i_M \in \{1,\ldots,n\}$, with $\fe_i$ being the $i$-th unit vector. By multiplication of \eqref{eqn:EI-definition} with $\fP^T$, we verify that
\begin{align}
(\tilde \fr(\fu;\fp))_{i_j} = \fr(\fu;\fp))_{i_j}, \quad j=1,\ldots,M.
\end{align}
In this sense the approximation acts as an interpolation within the set of magic points.

The identification of the interpolation 
points is an incremental procedure which is performed during the 
offline phase. We assume the existence of a set of training snapshots 
$\cY:=\{\fy_1,$ $\ldots,$ $\fy_{n_{\rm train}}\} \subset \ffR^n$ with 
${\mathrm{dim}(\mathrm{span}(\cY))\geq M}$.

Then a POD of these snapshots results in the 
collateral basis vectors $\fu_1,\ldots, \fu_M$ and we define  
$\fU_l := [\fu_1,\ldots, \fu_l]$ for $l=1,\ldots, M$.
The algorithm for the point selection is initialized with $\fP_0={[\,]}$, $I_0 := \emptyset$,
$\fU_0 = {[\,]}$ and then computes for $l=1\ldots, M$
\begin{align}
\fq_l  & := \fu_l - \fU_{l-1} (\fP_{l-1}^T \fU_{l-1})^{-1} \fP_{l-1}^T \fu_l, \\
i_l    & := \mathrm{arg}\max_{i\in \{1,\ldots,n\}} |(\fq_l)_i|, \\
\fP_l  & := [\fP_{l-1}, \fe_{i_l}], I_l := I_{l-1} \cup \{i_l\}. 
\end{align}
Finally, we set $\fP := \fP_M$, $\fU := \fU_M$ and $I := I_M$, which concludes the construction of $\tilde \fr$. 
Intuitively, in each iteration, 
the interpolation error $\fq_l$ for the current POD basis vector $\fu_l$ is determined, 
the vector entry $i_l$ with maximum absolute 
value is identified which gives the next index.
Regularity of the matrix $\fP^T \fU$ is required for a well-defined interpolation. 
This condition is automatically satisfied under the afore-mentioned assumption of a sufficiently 
rich set of snapshots $\cY$. As training set $\cY$, one can either use 
samples of the nonlinearity (\cite{DHO12}), POD-modes of these (\cite{CS12}), or use snapshots 
of the state vector or combinations thereof. In contrast to the instationary case, we may not use only training snapshots 
of $\fr$: As the residual $\fr$ is zero for all snapshots, we would try to find a basis for a 
zero-dimensional space $\mathrm{span}(\cY) = 0$, which is not possible for $M> 0$.
But the residual at the intermediate (non-equilibrium) iterates is non-zero and this is also a good target quantity 
for the (D)EIM, as these terms appear at the right hand side of the linear system during the 
fixed point iteration. Hence, a reasonable set $\cY$ is obtained via
\begin{align}
\cY = [ \fy_1, \dots, \fy_{n_{\rm train}} ] & = 
 [ \fr(\fu^{(0)}(\fp^{(1)}); \fp^{(1)}), \dots,   \fr(\fu^{(\alpha_1)}(\fp^{(1)}); \fp^{(1)}), \nonumber\\
 & \qquad \dots, \fr(\fu^{(\alpha_s)} (\fp^{(s)}); \fp^{(s)} ) ]
\end{align}
where $\alpha_1$, \dots, $\alpha_s$ are the number of fixed point iterations of the 
full simulation scheme for parameters $\fp^{(1)},\ldots,\fp^{(s)}$.

Inserting $\tilde \fr$ from (\ref{eqn:EI-definition})
for the nonlinearity into the full problem and projection by left-multi\-plication with 
a weight matrix $\fW \in \ffR^{n\times m}$ yields the POD-DEIM reduced $m$-dimensional nonlinear system for 
the unknown $\fgamma$
\begin{align}
\fW^T \fU (\fP^T \fU)^{-1} \fP^T \fr({\bar{\fu}_*(\fp) + }\fV \fgamma; \fp) = 0.
\end{align}
This low-dimensional nonlinear problem is iteratively solved by a fixed point procedure, i.e.\ 
at the current approximation $\fgamma^{(\alpha)}$ we solve the linearized 
problem for $\delta \fgamma^{(\alpha)}$ 
\begin{align}
\fW^T \fU (\fP^T \fU)^{-1}\fP^T \fJ( {\bar{\fu}_*(\fp) + } \fV\fgamma^{(\alpha)}; \fp ) \fV \delta \fgamma^{(\alpha)} = 
   - \fW^T \fU (\fP^T \fU)^{-1} \fP^T \fr( {\bar{\fu}_*(\fp) + } \fV \fgamma^{(\alpha)};\fp) \label{eqn:EIM-reduced-system}
\end{align}
and set $\fgamma^{(\alpha+1)}:= \fgamma^{(\alpha)}+\delta \fgamma^{(\alpha)}$.
As in the previous sections, if $M<{m}$, this linear system cannot be solved uniquely. In that case, an alternative 
would be to solve a residual least-squares problem, similar to the GNAT-procedure, cf.\ \cite{CBF11}.
Note, that the assembly of this system does not involve any high-dimensional operations, as the product of the first four matrices 
on the left and right hand side can 
be precomputed as a small matrix $\fX:= \fW^T \fU (\fP^T \fU)^{-1}$. Then, the terms $\fP^T \fJ, \fP^T \fr$ also do not require any 
high-dimensional operations, as the multiplication with $\fP^T (\cdot) = (\cdot)_I$ corresponds to evaluation of the ``magic'' 
rows of the Jacobian and right hand side, respectively. 
Typically in discretized PDEs, these $M$ rows only depend on few $\bar M$ entries of the unknown variable (e.g.\ 
the DOFs related to neighboring elements). This number $\bar M$ is typically bounded by a certain multiple of $M$ due to 
regularity constraints on the mesh \cite{DHO12}.

In \eqref{eqn:EIM-reduced-system} it can be recognized, what is required for the collateral basis $\fU$: 
If $\fJ( \fV\fgamma; \fp )\fV \delta \fgamma \in \mathrm{colspan}(\fU)$ 
and $\fr(\fV \fgamma;\fp) \in \mathrm{colspan}(\fU)$ then we are exactly solving the Galerkin-POD reduced 
linearized system
\begin{align} 
\fW^T \fJ( \fV\fgamma; \fp ) \fV \delta \fgamma = - \fW^T \fr(\fV \fgamma; \fp). \label{eqn:Galerkin-POD-reduced-system}
\end{align}
Hence, this gives a guideline for an alternative reasonable choice of the training set $\cY$, namely  
consisting of snapshots of both (columns of) $\fJ$ or $\fJ \fV$ and $\fr$.

In the Galerkin projection case, one can choose $\fW=\fV$. This is the choice that 
we pursue in the experiments to make the procedure more similar to the other reduction approaches.
The offline phase of the DEIM is summarized in \algoref{algo:deim-offline} and an algorithm of the online 
phase is  provided in \algoref{algo:deim}.

\begin{algorithm}[h]
\SetSideCommentRight
\SetNoFillComment
\SetAlgoSkip{medskip}
\LinesNumbered
\SetKwInOut{Input}{Input}
\SetKwInOut{Output}{Output}
\SetArgSty{text}
\caption{offline phase of the Discrete Empirical Interpolation Method (DEIM)}
\label{algo:deim-offline}

\Input{ collateral basis functions $\fu_k$ (and related matrix $\fU$) from POD of snapshots $\cY$}
\Output{ sampling matrix $\fP$;  magic point index set $I :=\{i_1,\ldots,i_M\}$;\\
reduced basis $(\hat{\fq}_l)_{l=1,\dots, M}$ (optional)}

set $\fP_0 \leftarrow {[\,]}$, $\fU_0 = {[\,]}$, $I_0 := \emptyset$  \tcp*{initialization}
\For{$l=1\ldots, M$}{
  $\fU_l \leftarrow [\fu_1,\ldots \fu_l]$ \tcp*{truncated POD matrix}
  $\fq_l \leftarrow \fu_l - \fU_{l-1} (\fP_{l-1}^T \fU_{l-1})^{-1} \fP_{l-1}^T \fu_l$ \tcp*{interpolation residual}
  $i_l   \leftarrow \mathrm{arg}\max_{i\in \{1,\ldots,n\}} |(\fq_l)_i|$ \tcp*{maximum of residual}
  $\hat{\fq}_l \leftarrow \fq_l / \left(\fq_l\right)_{i_l}$ \tcp*{normalization}
  $\fP_l  \leftarrow [\fP_{l-1}, \fe_{i_l}], \,\, I_l := I_{l-1}\cup\{i_l\}$ \tcp*{extend projection/magic points}
}
set $\fP := \fP_M, I := I_M$.
\end{algorithm}

\begin{algorithm}[h]
\SetSideCommentRight
\SetNoFillComment
\SetAlgoSkip{medskip}
\LinesNumbered
\SetKwInOut{Input}{Input}
\SetKwInOut{Output}{Output}
\SetArgSty{text}
\caption{online phase of the Discrete Empirical Interpolation Method (DEIM)}
\label{algo:deim}

\Input{ parameters $\fp\in\cP$ reduced basis $\fV$, POD-DEIM sampling matrix 
      $\fX:= \fW^T \fU (\fP^T \fU)^{-1}$ and magic point index set $I$}
\Output{ reduced vector $\fgamma$ and nodal temperatures $\tilde{\fu}$ {\it (optional)}}\vskip2pt
\hrule\vskip2pt
set $\fgamma^{(0)}=\fO$; ${\tilde{\fu}^{(0)}= \bar{\fu}_\ast(\fp)}$; $\alpha=0$ \tcp*{initialization}
$\bar \fJ \leftarrow (\fJ(\tilde{\fu}\ba;\fp)\fV)_I$ \tcp*{$ c_{\rm Jac}$; evaluate $M$ rows of right-projected Jacobian}
$\bar \fr \leftarrow (\fr(\tilde{\fu}\ba;\fp))_I$        \tcp*{$c_{\rm rhs} $; evaluate $M$ rows of right hand side}
solve $\fX \bar \fJ \delta \fgamma^{(\alpha)} = - \fX \bar \fr $ \tcp*{$c_{\rm sol} $; fixpoint iter.\ for $\delta \fgamma^{(\alpha)} $}
update $\fgamma^{(\alpha+1)} \leftarrow \fgamma^{(\alpha)} + \delta \fgamma^{(\alpha)}$,
compute $\tilde{\fu}^{(\alpha+1)} \leftarrow \tilde{\fu}\ba + \fV {\delta\fgamma\ba}$ and set $\alpha\leftarrow\alpha+1$  \tcp*{$c_{\rm reloc}$; update}
converged ($\Vert\delta\tilde{u}\ba\Vert_{L^2(\varOmega)}=\Vert\delta\fgamma\ba\Vert_{l^2}<\epsilon_{\rm max}$)? $\to$ {\bf end}; \quad {\it else:} goto \NlSty{2}
\end{algorithm}


\subsection{Hyper-Reduction (HR)}
\label{sec:hyperreduction}

In order to improve the numerical efficiency, the Hyper-Reduction method \cite{Ryckelynck-2005} introduces a Reduced Integration Domain (RID) denoted $\varOmega_Z\subset\Omega$. The RID depends on the reduced basis. It is constructed by offline algebraic operations. The hyper-reduced equations are a Petrov-Galerkin formulation of the equilibrium equations, obtained by using truncated test functions having zero values outside the RID. The vector form of the reduced equations is similar to the one obtained by the Missing Point Estimation method \cite{AstridWeilandEtAl2004} proposed for the Finite Volume Method. The strength of the Hyper-Reduction is  its ability to reduce mechanical models in material science while keepingthe formulation of the constitutive equations unchanged \cite{Ryckelynck-2009}. The smaller the RID, the lower the computational complexity and the higher the approximation errors. These points have been developed in previous papers dealing with various mechanical problems (e.g. \cite{Ryckelynck-2010,Ryckelynck-2012}).

The offline procedure of the Hyper-Reduction method involves two steps.
The first step is the construction of the Reduced Integration Domain $\varOmega_Z$.  
For the present benchmark test the RID is the union of a subdomain denoted by $\varOmega^u$ generated from the reduced vector gradients $( \nabla \psi_k)_{k=1,...,m}$,  
and a domain denoted by $\varOmega^+$ corresponding to a set of neighboring elements to the previous subdomain. Usually in the Hyper-reduction method, the user can select an additional subdomain of $\Omega$ in order to extend the RID over a region of interest. This subdomain is denoted by $\Omega_{user}$. In the sequel,  to get small RIDs, $\Omega_{user}$ is empty.
The set $\varOmega^u$ 
consists of aggregated contributions $\varOmega^u_k,k=1,\ldots,m$  
from all the reduced vectors:
\begin{align}
	\varOmega_Z &:= \varOmega^{u} \cup \varOmega^+ \cup \Omega_{user}, &
	\varOmega^{u} &:= \cup_{k=1}^{m} \varOmega^{u}_k .
	\label{RID1}
\end{align}

To give the full details of the procedure, we introduce the domain partition in finite elements denoted 
$(\varOmega^{\rm e}_j \subset \varOmega)_{j=1,...n_{\rm el}}$: $\varOmega = \cup_{j=1}^{n_{\rm el}} \varOmega^{\rm e}_j$, 
where $n_{\rm el}$ is the number of elements in the mesh. The domain~$\varOmega^{u}_k$ is the element where the 
maximum $L^2(\varOmega)$ norm of reduced vectors $\nabla \widetilde{\psi}_k$ is reached. In \cite{Ryckelynck-2005},  $\nabla \widetilde{\psi}_k$ was set equal to $\nabla \psi_k$. Here, when applying the DEIM  to $(\nabla \psi_k)_{k=1,\dots,m}$, the interpolation residuals provide a new reduced basis $(\fq_k)_{k=1,\dots,m}$ (cf.\  
\algoref{algo:hr-offline}) related to temperature gradients. In this paper, $\nabla \widetilde{\psi}_k$ is the output reduced basis produced by the DEIM, when it is applied to $(\nabla \psi_k)_{k=1,\dots,m}$.
Other procedures, for the RID construction, are available in previous papers on hyper-reduction. 
The element selection reads for ${k=1, \dots, m}$:
\begin{align}
\varOmega^{u}_k & = \underset{ \varOmega^{\rm e}_j, j=1, \dots, n_{\rm el} }{\hbox{arg \: max}} \left\|  \nabla \widetilde{\psi}_k \right\|_{L^2( \varOmega^{\rm e}_j)} , \label{RID2}
\end{align} 
where $\| . \|_{L^2( \varOmega^{\rm e}_j)}$ is the $L^2$ norm restricted to the element $\varOmega^{\rm e}_j$. 
Several layers of surrounding elements denoted $\varOmega^+$  can be added to $\varOmega^{u}$. 

The second step of the offline Hyper-Reduction procedure is the generation of truncated test functions which are zero outside of the RID. The truncation operator $P_Z$ is defined for all $u_{h} \in \cV^{\rm h}_0$ by
\begin{align}
 P_Z(u_{h}) & := \sum_{i \in \mathcal{F}} \varphi_i \: u_{h,i},  \quad
 \mathcal{F} = \Bigl\{ i \in \{1, \dots, n\} \: \Big| \: \intop_{\varOmega \backslash \varOmega_Z} \varphi_i \, \varphi_i \, \d \varOmega = 0 \Bigr\}
\end{align}
Here, $\mathcal{F}$ is the set of indices of internal points, i.e., inner FE nodes, in $\varOmega_Z$ which are related to the available FE equilibrium for predictions which are forecasted only over $\varOmega_Z$, i.e. for all $w \in \cV_0^h$ holds with $\varGamma_2^Z := \varGamma_2 \cap \partial \varOmega_Z$
\begin{equation}
a(u_h, P_Z (w); \fp) - l(P_Z (w)) =\! \intop_{\varOmega_Z} \mu(u_{h}; \fp) \, \nabla  u_{h}  \cdot \nabla (P_Z (w)) \d \varOmega  - \!\intop_{\varGamma_2^Z} P_Z (w) q_* \d \varGamma.
\end{equation}
The operator~$P_Z$ can be represented by a truncated projector denoted~$\fZ$. More precisely, if $\cF=\{i_1,i_2,\ldots, i_l\}$ with ${l := \mathrm{card}(\cF)}$, then 
\begin{align}
\fZ & := [ \fe_{i_1}, \fe_{i_2}, \ldots, \fe_{i_l}] \in \ffR^{n \times l} ,& P_Z(u_{h}) & := \sum_{i =1}^n \varphi_i \: (\fZ \: \fZ^T \: \fu)_i
\end{align}
with $\fe_i \in \ffR^n$ the $i$-th unit vector. Therefore, the hyper-reduced form of the linearized prediction step is: for a given ${\fgamma}^{(\alpha)}$, find $\delta \fgamma^{(\alpha)}$ such that,
\begin{align}
 \fV^{T} \: \fZ \: \fZ^T \: \fJ( \fV\fgamma^{(\alpha)}; \fp ) \: \fV \: \delta \fgamma^{(\alpha)} & = -  \fV^{T} \: \fZ \: \fZ^T \: \fr( \fV {\fgamma^{(\alpha)}}; \fp)   \label{lin_HROM1}
\end{align}
where $\fJ$ is given by \eqref{eq:J:approx} and $\fgamma^{(\alpha+1)} := \fgamma^{(\alpha)} +  \delta \fgamma^{(\alpha)}$.

In addition to $\fZ$ we introduce also the operator $\bar{\fZ}\in\ffR^{n\times{\bar{l}}}$ that is a truncated projection operator onto the $\bar{l}$ points contained in the RID. In practice, the discrete unknowns are computed at these $\bar{l}\geq l$ points in order to compute the residual at the inner points $l$. Note that often $\bar{l}$ is significantly larger than $l$, especially if the RID consists of disconnected (scattered) regions.

The complexity of the products related to the fixed point operator~$\fJ$ at the left hand side term scale with $2 \zeta l m + 2 l m^2$ where $\zeta$ is the maximum number of non-zero entries per row of $\fJ$. For the right hand side the computational complexity is $2 l m$. For both products, the complexity reduction factors are $n / l$. To obtain a well-posed hyper-reduced problem one requires to fulfill the following condition $l \geq m$. If this condition is not fulfilled, the linear system of equations \eqref{lin_HROM1} is rank deficient. In case of rank deficiency, one has to add more surrounding elements to the RID. The closer $l$ to $m$, and the lower $m$, the less complex is the solution of the hyper-reduced formulation. The RID construction must generate a sufficiently large RID. If not, the convergence can be hampered, the number of correction steps can be increased and, moreover, the accuracy of the prediction can suffer. When $\varOmega_Z=\varOmega$ then $\fZ$ is the identity matrix and the hyper-reduced formulation coincides with the usual system obtained by the Galerkin projection. An a posteriori error estimator for hyper-reduced approximations has recently been proposed in \cite{ryckelynck2015} for generalized standard materials.

The offline phase of the hyper-reduction is summarized in \algoref{algo:hr-offline} and an algorithm of the online phase is  provided in \algoref{algo:hr}.

\begin{algorithm}[h]
\SetSideCommentRight
\SetNoFillComment
\SetAlgoSkip{medskip}
\LinesNumbered
\SetKwInOut{Input}{Input}
\SetKwInOut{Output}{Output}
\SetArgSty{text}
\caption{offline phase of the Hyper-Reduction (HR)}
\label{algo:hr-offline}

\Input{ POD basis functions $\psi_k$ (and related matrix $\fV$) from snapshot POD, global matrix $\fG \in \mathbb{R}^{(n_{\rm gp}D) \times n }$ ($D:$ space dimension) for the gradient of the FE ansatz functions, the number $\ell$ of additional layers of elements used to extend the RID and 
the user-defined subdomain $\Omega_{user}\subset\Omega$}
\Output{ reduced integration domain~$\varOmega_Z$; truncated identities $\fZ, \bar{\fZ}$
}\vskip2pt
\hrule\vskip2pt

set $\fU= \fG \fV$, $\widetilde{\fU}=[ \, ]$, $\fP=[ \, ]$ \tcp*{initialization gradient at all int. points from RB}
\tcp{{$\fu_k\in\ffR^{n_{\rm gp}D}$ is a column vector containing $\nabla\psi_k$ at all $n_{\rm gp}$ int. points}}
\For(\tcp*[f]{select interpolation points for mode gradients}){$l=1, \dots, m$}{
$\fq \leftarrow \fu_l - \widetilde{\fU} (\fP^T \widetilde{\fU})^{-1}\fP^T\fu_l$\tcp*{project onto current sampling sites}
$j \leftarrow \underset{i\in \{1,\ldots,n_{\rm gp}D\} }{\mathrm{arg\ max}} |q_i|$ \tcp*{pick gradient component with largest amplitude}
$\fP \leftarrow [\fP \: \fe_{j}]$, \quad
$\widetilde{\fU} \leftarrow [\fu_1 \ldots \fu_{l}]$
\tcp*{enrich projection and sampling set}
$\varOmega_Z \leftarrow \varOmega_Z \cup $ \text{ElementContaining}$(j)$
}
\For(\tcp*[f]{add element layers}){$k=1, \dots, \ell$}{
$\varOmega_Z \leftarrow $ GrowOneElementLayer($\varOmega_Z$)}
$\varOmega_Z \leftarrow \varOmega_Z \cup \varOmega_{user}$ \tcp*{zone of interest is considered}
construct $\bar{\fZ}$ and $\fZ$ based on all/internal-only points of $\varOmega_Z$
\end{algorithm}

\begin{algorithm}[h]
\SetSideCommentRight
\SetNoFillComment
\SetAlgoSkip{medskip}
\LinesNumbered
\SetKwInOut{Input}{Input}
\SetKwInOut{Output}{Output}
\SetArgSty{text}
\caption{online phase of the Hyper-Reduction (HR)}
\label{algo:hr}

\Input{ parameter vector $\fp\in\cP$; scatter/gather operator $\fP_e$ of element $e$ and weight $v_i$ at the integration point $\fx_i$ $(i=1, \dots, n_{\rm gp})$}
\Output{ reduced vector $\fgamma(\fp)$ and nodal temperatures $\tilde{\fu}$ {\it (optional)}}\vskip2pt
\hrule\vskip2pt

set $\fu^{(0)}=\bar{\fu}_\ast(\fp)$; $\alpha=0$ \tcp*{initialization}
set $\fr=\fO$, $\fJ=\fO$   \tcp*{reset r.h.s. and Jacobian}
\For(\tcp*[f]{loop over reduced integration domain}){$e \in \varOmega_Z$}{
$\fr_e=\fO, \fK_e=\fO$\tcp*{initialize element residual and stiffness}
\For(\tcp*[f]{loop over the $n_e^{\rm gp}$ int. points of element $e$}){$i=1, \dots, n_e^{\rm gp}$}{
{evaluate the FE matrices $\fN, \fG$ and $\fG_e = \fG \fP_e $ at the current int. point}\tcp*{$c_{\rm reloc}$}
compute temperature $u_h\ba \leftarrow \fN\fu$ and gradient $\fg_h\ba \leftarrow \fG\fu\ba$ \tcp*{$c_{\rm reloc}$}
compute temperature $u_h\ba \leftarrow \fN \fu\ba { + \bar{u}_*(\fp)}$ and gradient $\fg_h\ba \leftarrow \fG\fu\ba$ at int. point \tcp*{$c_{\rm reloc}$}
evaluate constitutive model $\mu\ba \leftarrow \mu(u_h\ba; \fp)$ \tcp*{$c_{\rm const}$}
$\fr_e \leftarrow \fr_e + v_i \, \mu\ba \, {\fG_e}^T \fg_h\ba$ \tcp*{$c_{\rm rhs}$}
$\fJ_e \leftarrow \fJ_e + v_i \, \mu\ba \, {\fG_e}^T \fG_e $ \tcp*{$c_{\rm Jac}$}
}
$\fr \leftarrow \fr + \fP_e^T \fr_e$; $\fJ \leftarrow \fJ + \fP_e^T \fJ_e \fP_e$ \tcp*{$c_{\rm rhs}, c_{\rm Jac}$}
}
$\fr \leftarrow \fV^T \fZ \fZ^T \fr $  \tcp*{project residual at inner nodes of $\varOmega_Z$; $c_{\rm rhs}$}
$\fJ \leftarrow \fV^T \fZ \fZ^T \fJ \fV $  \tcp*{project Jacobian at inner nodes of $\varOmega_Z$; $c_{\rm Jac}$}
solve $\fJ\delta\fgamma\ba=-\fr$ \tcp*{$c_{\rm sol}$}
compute $\fu^{(\alpha+1)} \leftarrow \fu\ba + \fV\delta\fgamma\ba$ and set $\alpha\leftarrow\alpha+1$  \tcp*{update nodal temperatures; $c_{\rm reloc}$}
converged ($\Vert\delta\fgamma\ba\Vert_{l^2}<\epsilon_{\rm max}$)? $\to$ {\bf end}; \quad {\it else:} goto \NlSty{2}

\end{algorithm}

\subsection{Methodological comparison}
\label{sec:methodological:comparison}

We comment on some formal commonalities and differences between the HR and the (D)EIM.

We first note, that both methods reproduce the Galerkin-POD case, if $l=M=n$. For the HR this means 
that the RID is the full domain which implies that $\fZ$ is a square permutation matrix, hence being invertible and 
yielding $\fZ \fZ^T = \fI$, thus (\ref{lin_HROM1}) reduces to the POD-Galerkin reduced system  \eqref{eq:fixed:point:2}. 
For the (D)EIM this implies that the magic points consist of all grid points.
We similarly obtain that $\fP$ and $\fU$ are invertible and thus
$\fU (\fP^T \fU)^{-1} \fP^T = \fI$ and \eqref{eqn:EIM-reduced-system} also reproduces the POD-Galerkin 
reduced system \eqref{eq:fixed:point:2}. 

Further, we can state an equivalence of the DEIM and the HR under certain conditions, more precisely,  the reduced system of the HR is a special case of the DEIM reduced system. Let us assume that the sampling matrices coincide and the collateral basis also chosen as this sampling matrix, i.e. $\fU=\fP=\fZ$. Let us further assume that we have a Galerkin projection by choosing $\fW =\fV$ for the DEIM. Then $\fP^T \fU = \fZ^T \fZ = \fI_M$ is the $M$-dimensional identity matrix, hence we obtain
\begin{align}
  \fU (\fP^T \fU)^{-1} \fP^T  &= \fZ (\fZ^T \fZ )^{-1} \fZ^T = \fZ \fZ^T.
\end{align}
Then \eqref{eqn:EIM-reduced-system} yields
		\begin{align}
			\fV^T \fZ \fZ^T \fJ \fV \delta\fgamma &= - \fV^T \fZ \fZ^T \fr
		\end{align}
which exactly is the HR reduced system \eqref{lin_HROM1}.

A common aspect of HR, and (D)EIM obviously is the point selection by a sampling matrix.
The difference, however is the selection criterion of the internal points. 
In case of the DEIM these points are used as interpolation points, while for the HR they are used to specify
the reduced integration domain. 

A main difference of (D)EIM to HR is the way an additional collateral reduced-space is introduced in the reduced setting of the equations.
The HR is more simple by not using an additional basis related to the residuals, but the implicit assumption, that 
$\mathrm{colspan}(\fV)$ (which approximates $\fu$) also approximates $\fr$ and $\fJ$ well. This is a 
very reasonable assumption in symmetric elliptic problems and - in a certain way - it mimics the idea of having the same ansatz and test space as in any Galerkin formulation. But, from a mathematical point of view, it may not be valid in some more general cases, as in principle $\fU$ and $\fV$ are completely independent. E.g. we can multiply the vectorial residual 
equation \eqref{eq:residual:vec} by an arbitrary regular matrix, hence arbitrarily change $\fr$ (and thus $\fU$ for the DEIM), but not changing $\fu$ at all (i.e. not changing the POD-basis $\fU$). 
Hence, the collateral basis in the (D)EIM is first an additional technical ingredient and difficulty, 
which in turn allows to adopt the approximation space to the quantities, that need to be approximated well.

Theoretically, the EIM is well founded by analytical convergence results
\cite{MNPP07}.
But also, as a downside, the Lebesgue-constant, which essentially bounds the interpolation 
error to the best-approximation error, can grow exponentially.
The DEIM is substantiated with a-priori error estimates \cite{CS12}. 
We are not aware of such a-priori results for the HR, but also a-posteriori error control 
has been presented in \cite{ryckelynck2015}.


\subsection{Computational complexity}
\label{sec:cost}
With respect to the runtimes, we have decided not to provide numbers as these would heavily depend on the chosen implementation. In order to be more specific, the computational effort can generally be decomposed into:
\begin{itemize}\itemsep=0pt
  \item the computation of the local unknowns and of their gradients $c_{\rm reloc}$ (gradient/temperature computation),
  \item the evaluations of the (nonlinear) constitutive model $c_{\rm const}$,
  \item the assembly of the residual $c_{\rm rhs}$ and of the Jacobian $c_{\rm Jac}$, 
  \item the solution of the (dense) reduced linear system $c_{\rm sol}$.
\end{itemize}
The presented methods differ with respect to $c_{\rm reloc}$, $c_{\rm const}$, $c_{\rm rhs}$ and $c_{\rm Jac}$:
\begin{itemize}
  \item {\bf Finite Element simulation} ($n_{\rm gp}$: number of integration points; $n_{\rm el}$: number of elements; $n_{\rm el}^{\rm DOF}$: degrees of freedom per element)
  \begin{align*}
    c_{\rm reloc} & = 2 n_{\rm gp}  n_{\rm el}^{\rm DOF} && (\text{gradient/temperature computation})\\
    c_{\rm const} & \sim n_{\rm gp} && (\text{constitutive model})\\
    c_{\rm rhs} & = 2 n_{\rm gp} n_{\rm el}^{\rm DOF} && (\text{residual assembly})\\
    c_{\rm Jac} & = n_{\rm gp} \left[ (n_{\rm el}^{\rm DOF})^2  + 4 n_{\rm el}^{\rm DOF} \right] && (\text{Jacobian assembly}) \\
    c_{\rm sol} & \sim n^2.
  \end{align*}
  \item {\bf Galerkin-POD} 
  \begin{align*}
    c_{\rm reloc} & = 3 n_{\rm gp} m && (\text{gradient/temperature computation})\\
    c_{\rm const} & \sim n_{\rm gp} && (\text{constitutive model})\\
    c_{\rm rhs} & = 2 m n_{\rm gp} && (\text{direct residual assembly})\\
    c_{\rm Jac} & = (4 m + m^2) n_{\rm gp} && (\text{direct Jacobian assembly})\\
    c_{\rm sol} & \sim m^3.
  \end{align*}
\item {\bf Hyper-Reduction}\\
In the following $n^{\rm RID}_{\rm gp}$ is the number of integration points in the RID. Further, $c^{N,B}_{\rm FE}$ denotes the cost for the evaluation of $u$ and $\nabla u$ using the FE matrices $\fN$ and $\fG$ and $c_{\rm FE}^{r}$ is the related to the cost for the residual computation on element level (both at least linear in the number of nodes per element + scattered assembly + overhead) and $c^{K}_{\rm FE}$ is the cost related to the contribution to the element stiffness at one element ($\sim$number of nodes per element squared + scattered assembly + overhead). Lastly, $c_{\rm A}$ is the cost for the Jacobian assembly (i.e. matrix scatter operations).
  \begin{align*}
    c_{\rm reloc} & = \bar{l} m + n^{\rm RID}_{\rm gp} c^{N,B}_{\rm FE} && (\text{get }u, \partial_x u, \partial_y u\text{ in }\varOmega_Z)\\
    c_{\rm const} & \sim n^{\rm RID}_{\rm gp} && (\text{constitutive model})\\
    c_{\rm rhs} & = n^{\rm RID}_{\rm gp} c_{\rm FE}^{r} + m l && (\text{residual assembly and projection})\\
    c_{\rm Jac} & = (m\omega + m^2)l + n^{\rm RID}_{\rm gp} c^{K}_{\rm FE} + c_{\rm A}&& (\text{Jacobian assembly and projection})\\
    c_{\rm sol} & \sim m^3.
  \end{align*}
  \item {\bf Discrete Empirical Interpolation Method}\\
  The computational cost for the DEIM is closely related to the one of the HR by substituting $M$ for $l$ and $\bar{M}$ for 
  $\bar{l}$ (denoting the number of nodes which are needed to evaluate the residual at the $M$ magic points).
  In the cost notation of the \algoref{algo:deim}, we obtain 
  \begin{align*}
    c_{\rm reloc} & \sim m + n m\\
    c_{\rm rhs} & \sim M \bar M && (\text{residual assembly and projection})\\
    c_{\rm Jac} & \sim M \bar M m && (\text{Jacobian assembly and projection})\\
    c_{\rm sol} & \sim m^3.
  \end{align*}

\end{itemize}
From these considerations, the following conclusions can be drawn:
First, the number of integration points ($n_{\rm gp}$, $n^{\rm RID}_{\rm gp}$, $n^{\rm DEIM}_{\rm gp}$) required for the residual and Jacobian evaluation enter linearly into the effort.
Second, the reduced basis dimension~$m$ enters linearly into the residual assembly and both linearly and quadratically into the Jacobian assembly.
Third, for the HR and the (D)EIM, the ratio $\bar{l}/l$ and $\bar{M}/M$ have a significant impact on the efficiency: for the considered 2D problem with quadratic ansatz functions these ratios can range from 1 up to 21, i.e. for the same number of magic points pronounced variations in the runtime are in fact possible. The ratio $\bar{l}/l$ is determined by the topology of the RID, i.e. for connected RIDs it is smaller than for a scatter RID (i.e. for many disconnected regions forming the RID). Similarly, the (D)EIM has much smaller computational complexity in the case of magic points belonging to connected elements.
Fourth, the Galerkin-POD can be based on simplified algebraic operations as no nodal variables need to be computed. This is due to the fact that the reduced residual and Jacobian are directly assembled without recurse to nodal coordinates and to any standard FE routine.

\section{Numerical results}
\label{sec:experiments}
\subsection{ONLINE/OFFLINE decomposition and RB identification}
\begin{figure}[h]
  \centering
  \includegraphics[scale=1.2]{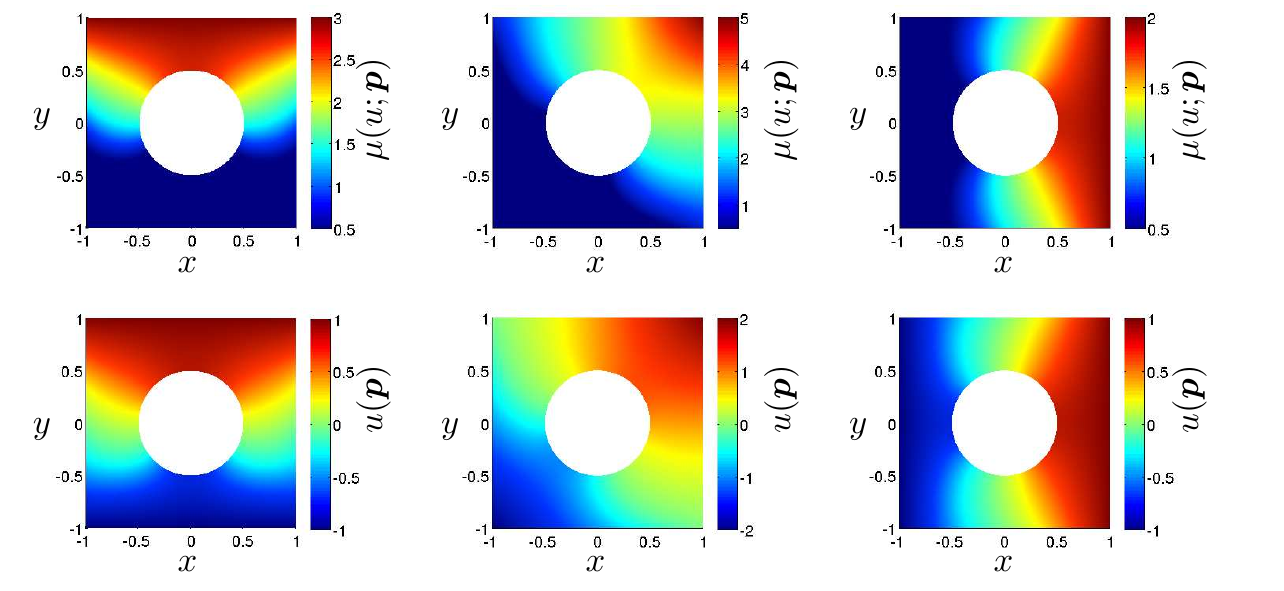}
  \caption{Parameter dependent conductivity $\mu(u; \fp)$ (top row) and solution $u(\fx; \fp)$ (bottom row) for three different snapshot parameters}
  \label{fig:snapshot:solutions}
\end{figure}

In the following we investigate the behavior of the heat conduction problem  \eqref{eq:heatproblem} for parameters which we recall from equation~(\ref{eq:param:range})
\begin{align}
  \fp & = [ g_{\rm x}, g_{\rm y}, c, \mu_0, \mu_1 ] \in [0,1] \times [0,1] \times [1, 2] \times \{ 1 \} \times \{0.5\}.
\end{align}
First, a regular parameter grid containing 125 equidistant snapshot points is generated and the high-fidelity FE model is solved for all those points yielding solutions $u_{h,i},i=1,\ldots,125$.
Here, the FEM discretization is based on a discretization into 800 biquadratic quadrilateral elements comprising a total of 2560 nodes (including 160 boundary nodes). The problem hence has $n=2400$ unknowns.
In order to exemplify the nonlinearity of the problem due to the temperature dependent conductivity, the conductivity (top row)  and the temperature field (bottom row) are shown for three different parameters in Figure~\ref{fig:snapshot:solutions}.

Then a snapshot POD is performed in order to obtain a RB from the snapshots. Different dimensions of the RB are considered with the approximation error
\begin{align}
  E_m &= \sqrt{ \frac{\sum_{i=1}^{125} \Vert {u_{h}(\fp_i) - \tilde{P} u_{h}(\fp_i)} \Vert^2_{L^2(\varOmega)}}{ \sum_{i=1}^{125} \Vert {u_{h}(\fp_i)} \Vert^2_{L^2(\varOmega)} } }
\end{align}
given in Table~\ref{tab:PODerror}. Here $\tilde{P}$ denotes the orthogonal projection operator onto the $m$ dimensional RB with respect to the standard inner product $\langle \cdot, \cdot \rangle_{L^2(\varOmega)}$ in the $L^2(\varOmega)$ function space
\begin{align}
\tilde{P}u_h(\fp) & = \sum_{i=1}^m \sum_{k=1}^m \psi_i \, ({\tilde{M}}^{-1})_{ik} \, \left\langle u_h(\fp), \psi_k \right\rangle_{L^2(\varOmega)}, &
{\tilde{M}_{ik}} &= \left\langle \psi_i, \psi_k \right\rangle_{L^2(\varOmega)} {\quad i, j = 1, \dots, m}.
\end{align}

\begin{table}[h]
  \centering
  \caption{Dimension of the RB vs. projection error for the snapshots}
  \label{tab:PODerror}
  \begin{tabular}{|l|c|c|c|c|c|}
   \hline
    dim. of RB & 16 & 24 & 32 & 48 & 60 \\ \hline
    $E_m$ & 1.107$\cdot$10\textsuperscript{-3} & 4.189$\cdot$10\textsuperscript{-4} & 2.583$\cdot$10\textsuperscript{-4} & 1.041$\cdot$10\textsuperscript{-4} & 5.661$\cdot$10\textsuperscript{-5} \\ \hline
  \end{tabular}
\end{table}

The low approximation errors given in Table~\ref{tab:PODerror} indicate, that the training data can essentially be represented rather accurately by the RB using a projection. Additionally, the projection error naturally decreases with increasing dimension. However, the solution of the reduced problem does not necessarily follow the same monotonicity. Since $\mu$ is bounded away from 0 due to $\mu_0>0$ the problem under consideration is coercive.
Similar to the linear case we expect that the approximation error $e(\fp) = {\tilde{w}(\fgamma(\fp);\fp) - w_{h}(\fp)}$ and the projection error are comparable in the sense that
\begin{align}
{\eta}(\fp) := \frac{  \Vert e(\fp) \Vert_{L^2(\varOmega)}  
}{\Vert u_h - \tilde{P} u_h \Vert_{L^2(\varOmega)}}
\label{eqn:Cdef}
\end{align}
is small. Due to the best-approximation by the orthogonal projection, we certainly  
have ${\eta}(\fp) \geq 1$.
The constant ${\eta}(\fp)$ cannot be provided in closed form for the considered nonlinear problem, but it can 
only be computed a posteriori by means of comparing the reduced to the full solution. Numerical 
values are provided in the following.

\subsection{Test cases}
In order to investigate the accuracy of the reduced models additional parameter vectors~$\fp$ need to be considered that are not matching the training data. Two test cases are considered in the sequel:
\begin{enumerate}
  \item[\bf{[}A{]}] A diagonal in the parameter space is considered with
  \begin{align}
    \fp\bj & := \fp_0 + \beta\bj ( \hat{\fp} - \fp_0 ), &
    \fp_0 & := [ 0, 0, 1, 1, 1/2 ], &
    \hat{\fp} & := [ 1, 1, 2, 1, 1/2 ] .
  \end{align}
  A total 101 equally spaced values of $\beta\bj$ was chosen, i.e. $\beta\bj= \frac{j}{100}$ for $j=0, 1,\dots, 100$.
  \item[\bf{[}B{]}] A set of 1000 random parameters $\fp\bj$ was generated using a uniform distribution in parameter space, i.e. a uniform distribution $\cU([0, 1])$ was chosen for $g_{\rm x}$, $g_{\rm y}$ and the parameter $c$ was assumed to be distributed via $\cU([1, 2])$.
\end{enumerate}

\subsection{Certification of the Galerkin RB method}
First, the ability of the Galerkin RB solution to approximate the optimal orthogonal projection and, thereby, the high-fidelity 
solution, was verified. Therefore, the constant ${\eta}(\fp)$ was evaluated for all snapshots of case {\bf[A]} and {\bf[B]}. The 
minimum, the mean and the maximum of ${\eta}(\fp)$ were determined for the 101 and 1000 test of case {\bf[A]} and {\bf[B]}, respectively. 
The results 
shown in Table~\ref{tab:POD:vs:PROJ} state the POD approximation error is found close to the projection error. This confirms 
the quality of the chosen RB. 

\begin{table}[h]
  \centering
  \caption{Computed values of ${\eta}(\fp)$ for different modes sets and for test cases {\bf[A]}, {\bf[B]}}
  \label{tab:POD:vs:PROJ}
  \begin{tabular}{|l|c|c|c|c|c|c|}
  \hline
  test case & \multicolumn{3}{c|}{\bf [A]} & \multicolumn{3}{c|}{\bf [B]} \\ \hline
  & min. & mean & max. & min. & mean & max. \\ \hline
  $m=16$ & 1.000 & 1.4708 & 1.8248 & 1.000 & 1.3185 & 2.0239 \\ \hline
  $m=24$ & 1.000 & 1.9088 & 2.8926 & 1.000 & 1.3187 & 2.6559 \\ \hline
  $m=32$ & 1.000 & 1.7273 & 2.7441 & 1.000 & 1.3679 & 2.4393 \\ \hline
  $m=48$ & 1.000 & 1.5386 & 2.0232 & 1.000 & 1.3051 & 1.9371 \\ \hline
  $m=60$ & 1.000 & 1.5096 & 1.9333 & 1.000 & 1.3447 & 1.8922 \\ \hline
  \end{tabular}
\end{table}
Note that in test case {\bf [B]} only a finite number of random parameter vectors was chosen which does not necessarily contain the extreme values of ${\eta}(\fp)$. The numerical data in Table~\ref{tab:POD:vs:PROJ} for test case {\bf [A]} shows that indeed, {\bf [A]} contains 
parameters leading to larger values of ${\eta}(\fp)$. When increasing the size of the random parameter set for {\bf[B]}, the maximum values of ${\eta}(\fp)$ in case {\bf[B]} should be equal or larger than the maximum values of case {\bf[A]}.

In addition to the error magnification parameter ${\eta}(\fp)$, the minimum, average and maximum of the relative error
\begin{align}
  \delta( \tilde{u}(\fp), \fp ) &= \frac{ \Vert \tilde{u}(\fp) - u_h(\fp) \Vert_{L^2(\varOmega)} }{ \Vert u_h(\fp) \Vert_{L^2(\varOmega)} }
\end{align}
are also computed for all samples. The results are provided in Table~\ref{tab:POD:relerror}. Note that for test case {\bf[A]} the 
minimum error is truly zero for $g_{\rm z}=g_{\rm y}=0$ which implies a homogeneous zero temperature. For a RB of dimension 32 the 
mean error is well below 10\textsuperscript{-3} for all tests and the maximum error over all tests is 
3.23$\cdot$10\textsuperscript{-3}. This basis provides a compromise between accuracy and computational cost and is therefore used 
for the comparison of the methods in the sequel.

\begin{table}[h]
  \centering
  \caption{Relative error of the Galerkin RB approximation}
  \label{tab:POD:relerror}
  \begin{tabular}{|l|c|c|c|c|c|c|}
  \hline
  test case & \multicolumn{3}{c|}{\bf [A]} & \multicolumn{3}{c|}{\bf [B]} \\ \hline
  & min. & mean & max. & min. & mean & max. \\ \hline
  $m=16$ & 
0.00 \, 10\textsuperscript{-3} & 2.54 \, 10\textsuperscript{-3} & 7.28 \, 10\textsuperscript{-3}
& 0.67 \, 10\textsuperscript{-3} & 1.75 \, 10\textsuperscript{-3} & 7.49\, 10\textsuperscript{-3} \\ \hline
  $m=24$ & 0.00 \, 10\textsuperscript{-3} & 1.07 \, 10\textsuperscript{-3} & 3.80 \, 10\textsuperscript{-3} 
  & 0.16 \, 10\textsuperscript{-3} & 0.88 \, 10\textsuperscript{-3} & 5.29\, 10\textsuperscript{-3}\\ \hline
  $m=32$ & 0.00 \, 10\textsuperscript{-3} & 0.82 \, 10\textsuperscript{-3} & 3.23\, 10\textsuperscript{-3}
  & 0.18 \, 10\textsuperscript{-3} & 0.63 \, 10\textsuperscript{-3} & 3.13\, 10\textsuperscript{-3} \\ \hline
  $m=48$ & 0.00 \, 10\textsuperscript{-3} & 0.31 \, 10\textsuperscript{-3} & 1.14\, 10\textsuperscript{-3} 
  & 0.06 \, 10\textsuperscript{-3} & 0.33 \, 10\textsuperscript{-3} & 2.07\, 10\textsuperscript{-3}\\ \hline
  $m=60$ & 0.00\, 10\textsuperscript{-3} & 0.20 \, 10\textsuperscript{-3} & 0.70\, 10\textsuperscript{-3} 
  & 0.05 \, 10\textsuperscript{-3} & 0.26 \, 10\textsuperscript{-3} & 1.55\, 10\textsuperscript{-3}\\ \hline
  \end{tabular}
\end{table}
Note that the slow decay of the accuracy of the Galerkin approximation indicated by the data provided in Table~\ref{tab:POD:relerror} indicates that the information content captured in the training snapshots is not sufficient to provide better accuracies. Therefore we decided on a dimension $m=32$ for the reduced basis in the subsequent experiments.

With respect to the computational efficiency of the RB solution it is observed that, unfortunately, the Galerkin RB solution is 
almost as costly as the high fidelity solution. This is due to the fact that the repeated assembly of the reduced system is computationally 
more intense {than the actual solving (which is typical for Matlab implementations)}. 
This aspect becomes more important for rather low-dimensional problems such as the one considered here. In order to 
actually reduce the computational cost the use of additional reduction techniques such as the Hyper-Reduction (HR) and Discrete 
Empirical Interpolation Method (DEIM) is required. Note also that the reduction techniques using a POD basis are only 
approximations of the Galerkin RB. Hence, the HR and DEIM cannot be better than the Galerkin RB solution except in few 
cases where ${\eta}_{\rm HR}(\fp)<{\eta}(\fp)$ or ${\eta}_{\rm DEIM}(\fp)<{\eta}(\fp)$, where 
${\eta}_{\rm HR}(\fp)$ and ${\eta}_{\rm DEIM}(\fp)$ denote the constant ${\eta}$ from (\ref{eqn:Cdef}) for the HR and DEIM method.
In our numerical tests this occurred only exceptionally. {Let us now turn to a more realistic multi-query situation where RB is crucial for non-prohibitive runtimes.}


\subsection{Application to Uncertainty Quantification}
\label{sec:res:UQ}
In real world simulation scenarios, material coefficients and boundary conditions are often not exactly known and one is interested in the impact of this uncertainty on the quantities of interest. To this end, uncertainty quantification (UQ) has been proposed and has become an active research field on its own. In classical forward UQ, the critical parameters are modeled as random variables; the distributions and correlation are derived from measurements as for example shown for nonlinear material curves in \cite{Romer_2016aa}. Finally, the forward model is evaluated at collocation points $\fp_i$ in the parameter space according to a quadrature method as e.g. Monte Carlo, \cite{Xiu_2010aa}. Typically many collocations (or `sampling') points are needed and therefore model order reduction has been shown to significantly reduce the computational costs, e.g. \cite{Haasdonk_2013aa}.

In the case of our thermal benchmark problem, the parameter vector $\fp=\fP(\omega)$ is considered as a realization of the random vector, where $\omega\in\varOmega_{\mathrm p}$ and $(\varOmega_{\mathrm p},\mathcal{F},\mathcal{P})$ is the usual probability space. We refer to this as test case~{\bf[C]} and assume that the random variables are independent and uniformly distributed as already introduced in test case~{\bf[B]}
\begin{align*}
	\fP(\omega) &=[G_{\rm x}(\omega), G_{\rm y}(\omega),C(\omega)]
	\qquad\text{with}\quad
	G_{\rm x}, G_{\rm y}\sim\cU([0, 1])
	\quad\text{and}\quad
	C\sim\cU([1, 2]).
\end{align*}
Finally, the statistical moments of the solution $u_{\star}(\fx;\fP(\omega))$ are approximated by the Monte Carlo method, e.g. \cite{Xiu_2010aa}
\begin{align}
	\ffE \bigl( u_{\star}(\fx; \fP) \bigr)
	&\approx \frac{1}{n_\mathrm{p}} \sum_{j=1}^{n_\mathrm{p}} u_{\star}(\fx; \fp\bj) 
	=: \bar{u}_{\star}(\fx)\\
  \ffM^{k} \bigl( u_{\star}(\fx; \fP) \bigr)
	&\approx \frac{1}{n_\mathrm{p}} \sum_{j=1}^{n_\mathrm{p}}
		\Bigl(
			u_{\star}(\fx; \fp\bj)
			-
			\bar{u}_{\star}(\fx)
			\Bigr)^k
	=: m_{\star}^{k}\bigl(\fx\bigr)
\end{align}
where $\fp\bj$ are the same $n_\mathrm{p}=1000$ random sample points generated for case~{\bf[B]} and $u_{\star}$ and $m_{\star}^k$ are the approximations of the solution and its $k$-th centered moment ($k>1$) obtained from Finite Elements, Galerkin-RB, DEIM and HR, i.e., $\star \in \{\mathrm{h},\mathrm{RB},\mathrm{DEIM},\mathrm{HR}\}$, respectively. For the DEIM we selected $M=400$ magic points.

\begin{figure}
  \centering
  \includegraphics[height=70mm]{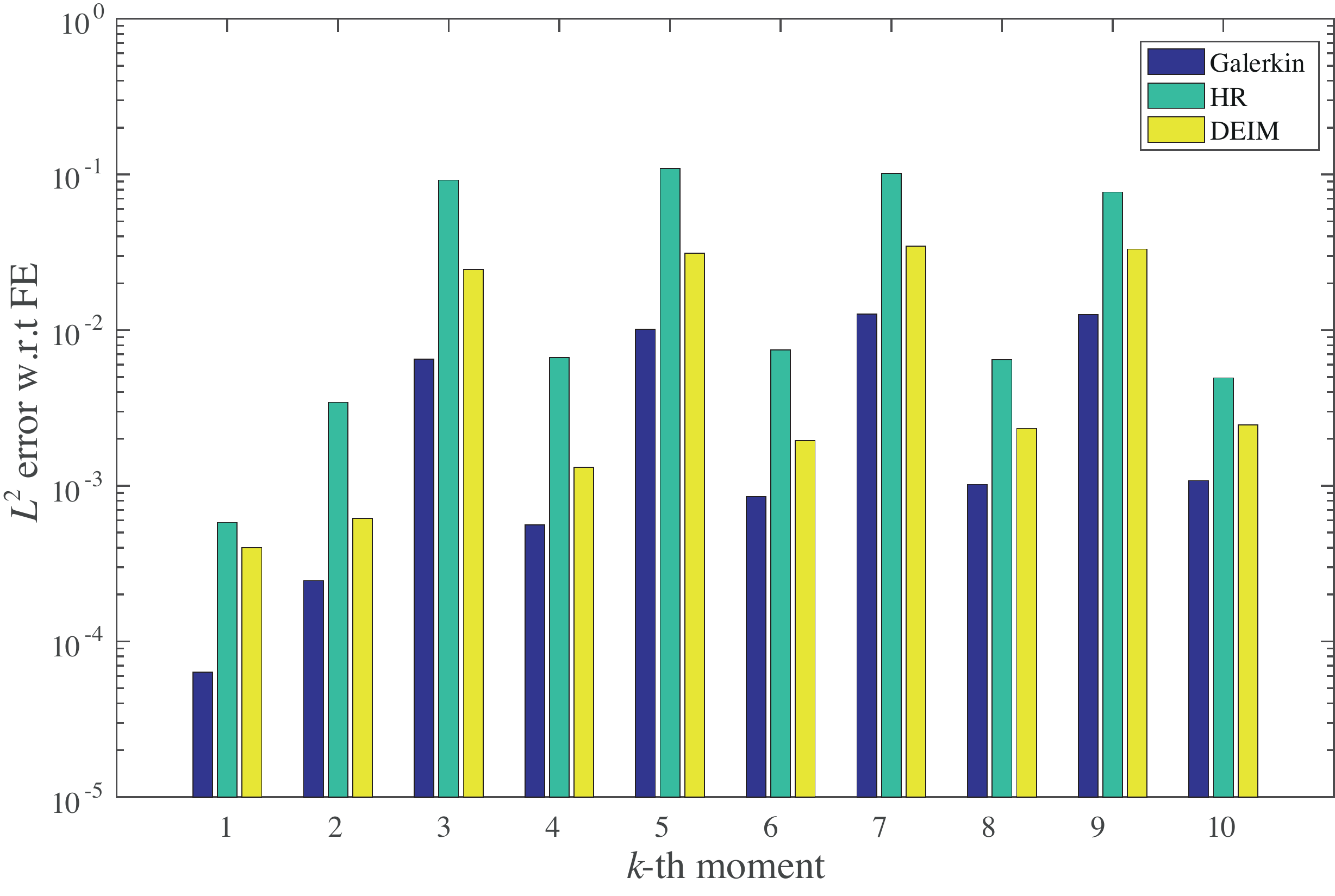}
  \caption{$L^2(\varOmega)$ error of the centered moments w.r.t. the finite element solution. Please note that we have set $m_{\star}^{1}:=\bar{u}_{\star}$ for simplicity of notation}
  \label{fig:uq_moments} 
\end{figure}

An estimation of the normalized root mean square error of the finite element solution due to Monte Carlo sampling can be obtained by  
\begin{align*}
	E_\mathrm{MC} &= \frac{1}{\sqrt{n_\mathrm{p}}\;\|\bar{u}_\mathrm{FE} \|_{L^2(\varOmega)} } \left\|\sqrt{m_h^{2}}\right\|_{L^2(\varOmega)}
	\approx 1.78\cdot10^{-2} \equiv 1.78\%.
\end{align*}
This implies that the accuracy of the reduced models in the prediction of~$\bar{u}_{\star}$ should be around 2\% or better, in order to render the reduced models capable of providing meaningful quantitative statistical information. Figure~\ref{fig:uq_moments} shows the error in the moments computed for the various reduction techniques with respect to the finite element reference solution
\begin{align*}
	E^k_\star &= \frac{\| m_{\star}^{k}-m_h^{k} \|_{L^2(\varOmega)} }{\|m_h^{k}\|_{L^2(\varOmega)}} .
\end{align*}

The figure indicates that the approximations of the expected value $\bar{u}_{\star}$ are at least accurate up to $10^{-3}$ and thus below (i.e. better than) the sampling accuracy $E_\mathrm{MC}$. Generally, the DEIM tends to perform better than the HR; the largest errors occur for $E^7_\mathrm{HR} =1.09\cdot10^{-1}$ and $E^7_\mathrm{DEIM} = 3.46\cdot10^{-2}$, respectively.

\subsection{Accuracy of the HR and DEIM vs. number of interpolation points}
Both, the HR and the DEIM sample the nonlinearity only at entries of the right hand side: the interpolation points. The HR has one 
additional parameter~$\ell$ describing how many element layers around a certain degree of freedom  (DOF) should be considered in order to generate $\varOmega^+$ which add additional interpolation points to the RID. 
The location of the DOF around which the interpolation points are located are selected based on the criterion described in Section~\ref{sec:hyperreduction}. In contrast to that the DEIM selects the points using only the right hand side information of the system. For the DEIM the number of sampling points $M$ is an input parameter describing the dimension of the collateral basis. The effect of the number of points is investigated in the following. Based on the considerations in Section~\ref{sec:methodological:comparison} both the HR and the DEIM should reproduce the POD solution for a large number of sampling sites while for a lower number of points the accuracy is trade-in for computational efficiency.

\begin{figure}[h]
  \centering
  \includegraphics[height=40mm]{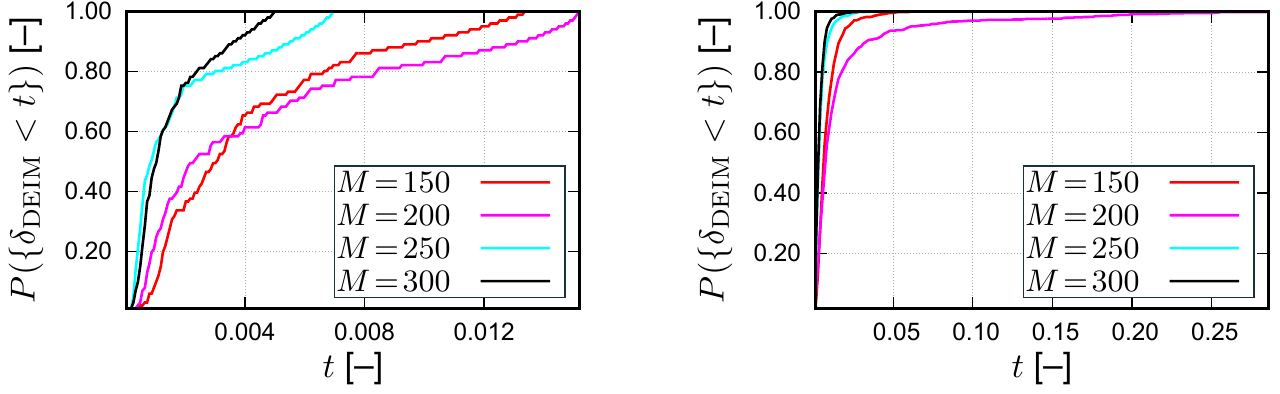}
  \begin{minipage}{0.49\textwidth}
    \centering
    {test case {\bf[A]}}
  \end{minipage}
  \begin{minipage}{0.49\textwidth}
    \centering
    {test case {\bf[B]}}
  \end{minipage}
  \caption{Statistical distribution function of the relative error of the DEIM for different numbers of magic points ${M \in \{150, 200, 250, 300 \}}$ (dimension of POD basis: $m=32$)}
  \label{fig:NMP:comparison}
\end{figure}

For the DEIM, different interpolation point numbers are considered for both test cases {\bf[A]} and {\bf[B]}. The resulting relative errors are compared in Figure~\ref{fig:NMP:comparison} in terms of the statistical distribution function $P(t)$ of the relative error, i.e., the probability of finding a relative error $\delta$ that is smaller or equal than $t$. Obviously the number of interpolation points has a pronounced impact on the distribution. Generally, the error function for low numbers of points states a significantly increase of the computational error due to DEIM in comparison with the POD. With increasing number of points the distribution function approaches the one of the POD. In our test the use of more than 300 sampling points can only improve the accuracy in a minor way. 
We must note that, in general, the accuracy of the DEIM must not be a monotonic function of interpolation points number. 

\begin{figure}
\centering
 \includegraphics[height=40mm]{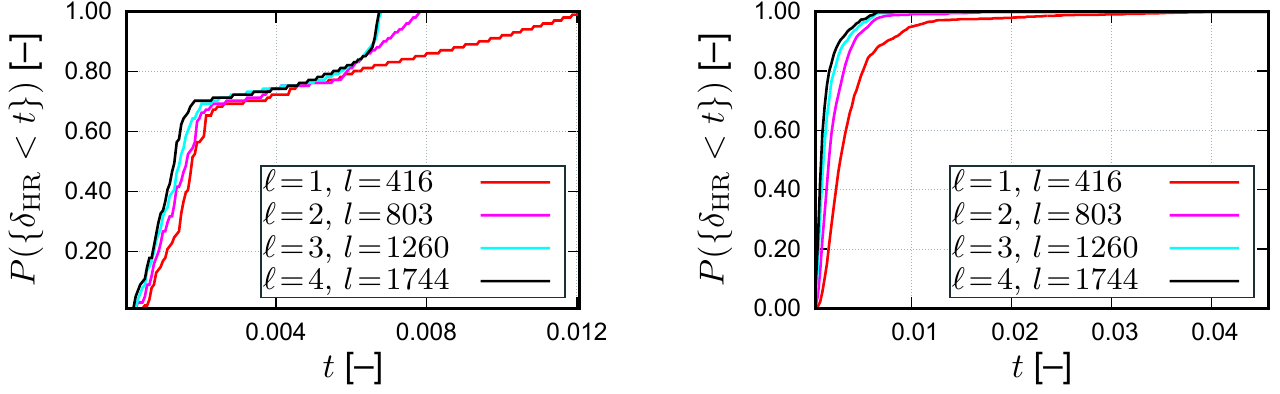}
  \begin{minipage}{0.49\textwidth}
    \centering
    {test case {\bf[A]}}
  \end{minipage}
  \begin{minipage}{0.49\textwidth}
  \centering
    {test case {\bf[B]}}
  \end{minipage}
  \caption{Statistical distribution function of the relative error of the hyper-reduction for different layer of elements added to the RID ${\ell \in \{1, 2, 3, 4 \}}$ (dimension of POD basis: $m=32$)}
  \label{fig:HR:comparison}
\end{figure}

For the hyper-reduced predictions, different layers of elements are added in $\varOmega^+$ in order to extend the RID. We have considered here $\ell=$ 1, 2, 3 and 4 layers of elements connected to $\varOmega^u$, for both test cases {\bf[A]} and {\bf[B]}. The resulting relative errors are compared in Figure~\ref{fig:HR:comparison} in terms of the statistical distribution function $P(t)$ of the relative error. With increasing number of layers the distribution function approaches the one of the POD. The number of internal points, $l$, increases rapidly when increasing the number of layers. In the case of the DEIM the growth of interpolation points is much more progressive. More than two layers of elements do not improve the accuracy significantly.

\begin{figure}
 \centering
 \includegraphics[height=45mm]{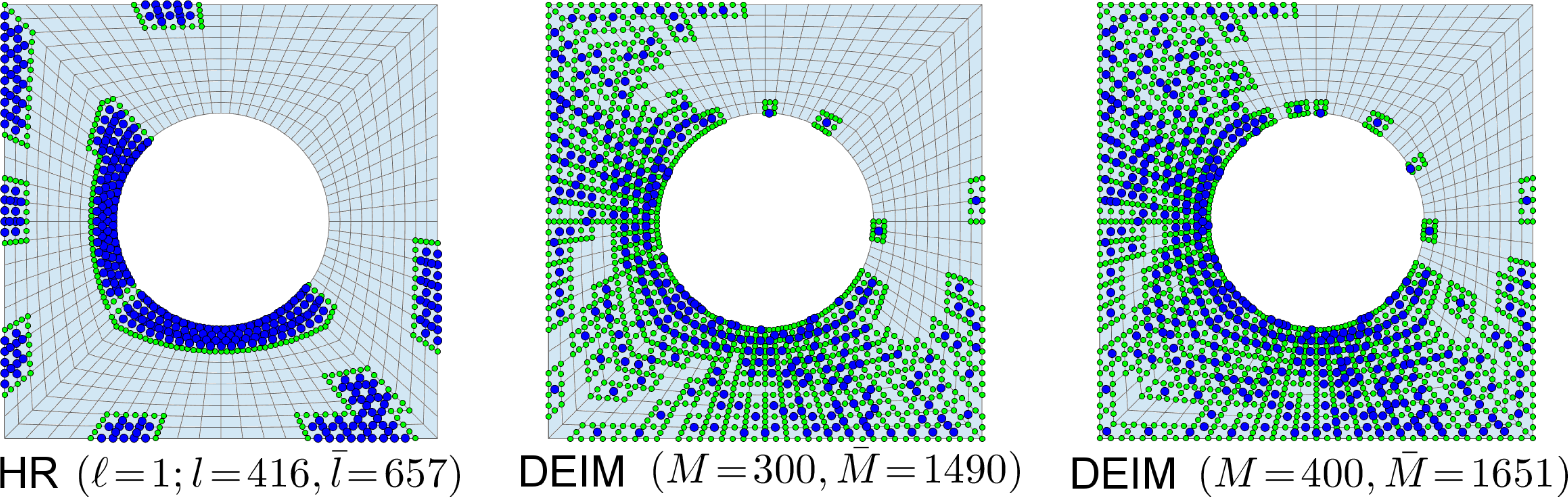}
 \caption{Position of the magic points (blue points) and the additional points required for the evaluation of the residual (green points); Hyper-reduction (left) vs. DEIM (middle, right) for $m=32$}
 \label{fig:magicpoints}
\end{figure}

One layer of additional elements gives predictions less accurate than the DEIM for approximately the same number of internal points/interpolation points (here: $l=416$ for the HR and  $M=300$ for the DEIM). However, the number of additional points required for the residual evaluation differs considerable: $\bar{l}=657$ vs. $\bar{M}=1490$. This can be explained by direct comparison of the reduced domains in Figure~\ref{fig:magicpoints}. For the Hyper-reduction the RID is rather compact (leftmost plot) while the magic points of the DEIM are rather scattered, thus requiring the temperature evaluation at many additional points.

The predictions of the Hyper-reduction are less accurate especially which can especially be seen by comparing Figure~\ref{fig:NMP:comparison} (left, test case {\bf[A]}, black line) and Figure~\ref{fig:HR:comparison} (left, test case {\bf[A]}, red line), where 80\% of the samples lead to errors below $\approx0.002$ for the DEIM and $\approx0.01$ for the HR. Nevertheless, the accuracy of the hyper-reduced predictions is generally of the same order of magnitude as the accuracy of the DEIM.

\section{Summary and conclusion}
\label{sec:conclusion}

The presented study revisits a - at first sight - rather simple nonlinear heat conduction problem. In order to accelerate the nonlinear computations, a Galerkin reduced basis ansatz is proposed (see Section~\ref{sec:galerkin}) using preliminary offline computations in the established framework of the snapshot POD. 

Both the HR and the (D)EIM can achieve significant accelerations of the computing time. These scale approximately with the number of magic points (here: $l$ or $M$) and/or with the number of points connected to the magic points ($\bar l$ and $\bar M$, respectively). In the presented examples less than 25\% of the original mesh were considered in both, the (D)EIM and the HR. By virtue of the considerations presented in Section~\ref{sec:cost} the computing times are reduced accordingly.

The selection of the magic points in the HR and the (D)EIM requires the computation of the solution at an increased number of nodes, i.e. $\bar{l}\geq l$ and $\bar{M} \geq M$, respectively. The higher the space dimension (2D, 3D, \dots), the more scattered the magic points and the higher the number of degrees of freedom per element, the more are $\bar{l}$ and $\bar{M}$ increased in comparison to $l$ and $M$.

Our examples also show that coarse predictions of the HR resulted in more robust computations than for the same number of magic points in the DEIM. However, the DEIM can achieve accuracies that are better than the ones of the HR in some situations, i.e. if the collateral basis is sufficiently accurate.

With respect to the implementation it shall be noted that the HR is less intrusive than the (D)EIM as it uses standard simulation outputs to generate the modes while the (D)EIM requires additional outputs for the construction of the collateral basis. Other than that, both techniques can be implemented using mostly the same implementation which is also confirmed by the similarity of both techniques presented in Section~\ref{sec:methodological:comparison}.

\section*{Acknowledgments}
The authors acknowledge the generous funding of the CoSiMOR scientific network by the German Research Foundation/Deutsche Forschungsgemeinschaft (DFG) under grants DFG-FR-2702/4-1, DFG-FR-2702/7-1. Felix Fritzen is thankful for financial support in the framework of the DFG Emmy-Noether group EMMA under grant DFG-FR2702/6. Sebastian Sch\"ops acknowledges support from the Excellence Initiative of the German Federal and State Governments and the Graduate School of CE at TU Darmstadt.

\end{document}